\documentclass[sigconf]{acmart}

\usepackage{subcaption}  
\usepackage{comment}
\usepackage{xcolor}
\usepackage{colortbl}
\usepackage{array}
\usepackage{enumitem}
\setlist{itemsep=2pt, parsep=0pt}

\usepackage{algorithm}
\usepackage{algpseudocode}  

\usepackage{multirow,booktabs}
\usepackage{tabularx}
\usepackage{caption}
\newtheorem{theorem}{Theorem}[section]

\newtheorem{proposition}[theorem]{Proposition}

\theoremstyle{definition}

\theoremstyle{remark}

\usepackage{mymacros}
\AtBeginDocument{%
  }

\setcopyright{none}
\acmConference[KDD '26, Workshop on Two-sided Marketplace Optimization (TSMO): Search, Discovery, Matching, Pricing \& Growth]{}{August 9-13}{Jeju, Korea}
\acmBooktitle{Workshop on Two-sided Marketplace Optimization (TSMO) at KDD '26, Jeju, Korea}
\acmDOI{}
\acmISBN{}
\makeatletter
\gdef\acmConference@date{}
\gdef\@copyrightyear{}
\makeatother

\begin{document}

\title{A Unified Approach to Interpretable Causal Root Cause Attribution}

\author{Jing Zhou}
\affiliation{%
  \institution{Amazon}
  \city{Tokyo}
  \country{Japan}}
\email{zhoujj@amazon.co.jp}

\author{Dominik Janzing}
\affiliation{%
  \institution{Amazon}
  \city{Tuebingen}
  \country{Germany}}
\email{janzind@amazon.de}

\author{Sepp Tsang}
\affiliation{%
  \institution{Amazon}
  \city{Tokyo}
  \country{Japan}}
\email{tsepp@amazon.co.jp}

\author{Patrick Blöbaum}
\affiliation{%
  \institution{Amazon}
  \city{Santa Clara}
  \country{United States}}
\email{bloebp@amazon.com}

\author{Marco Visentini Scarzanella}
\affiliation{%
  \institution{Amazon}
  \city{Tokyo}
  \country{Japan}}
\email{marcovs@amazon.co.jp}

\renewcommand{\shortauthors}{Zhou et al.}

\begin{abstract}
Understanding why marketplace metrics change is a central problem in two-sided marketplace optimization.
We study root cause attribution for metric changes in complex e-commerce systems, focusing on trade-offs between interpretability, efficiency, and causal validity. 
As a starting point, we extend a metric-decomposition method into a recursive metric-tree framework for multi-level root cause analysis, but this relies on independence and decomposability assumptions that miss complex causal dependencies. In contrast, graphical causal models (GCMs) relax these assumptions and improve causal validity, at the cost of interpretability, higher computational and data demands, and potential attribution target misalignment.
Through real-world applications, mathematical proofs, and simulations, we characterize the fundamental sources of these trade-offs. Guided by these insights, we propose a unified, causally informed attribution approach that integrates structural causal information into the metric-tree decomposition framework and corrects key sources of misalignment in GCM-based causal attributions, substantially improving causal validity while preserving interpretability and fast computation. Analytical proofs and simulations demonstrate that the proposed approach produces more accurate root cause attributions, and we also present a real-world application. 

\end{abstract}






\maketitle
\makeatletter
\fancyhead[LE]{\ACM@linecountL\@headfootfont\footnotesize KDD '26 Workshop, Jeju, Korea}
\fancyhead[RO]{\@headfootfont KDD '26 Workshop, Jeju, Korea\ACM@linecountR}
\makeatother


\section{Introduction} 
Two-sided marketplaces are operated through tightly coupled metrics spanning both the demand side (e.g., traffic, conversion, engagement) and the supply side (e.g., selection, price, availability). In practice, a recurring and high-stakes task on these platforms is understanding why key performance metrics change: for example, why did revenue drop this month, was the change driven mainly by traffic, pricing, or selection, and which levers should be prioritized next? 
These are not rare anomaly cases, but routine operational questions that directly affect growth, pricing, and seller-side strategy in large-scale online platforms.
 
A natural way to formalize such questions is as root cause attribution for metric changes across interacting marketplace variables.
Prior work on root cause analysis focuses primarily on anomaly or outlier detection \cite{li2022CIRCA,yan2022CMMD, nguyen2024outlier2, peiris2014pad,aksar2024runtime,sun2025ODE}.
However, a more relevant work \cite{GCM2021} framed root causing as a distribution change problem and used graphical causal models (GCMs, \cite{pearl2009causality}) to attribute distribution changes to local causal mechanisms. They achieved this by factorizing the joint distribution into causal mechanisms and using Shapley values \cite{Shapley1953,lundberg2017shap} to quantify each mechanism’s contribution to the overall shift.
But this approach suffers from interpretability, computational cost, and data demands, making it impractical for real-time marketplace monitoring.
There is also a long-standing decomposition literature that attributes differences or changes in aggregate quantities to underlying factors
\cite{oaxaca1973maleFemale,blinder1973wage, shorrocks1982inequality, ang2005LMDI}. However, these methods were not designed for high-frequency metrics monitoring in modern online systems, and hence, not as interpretable. 
In parallel, metric decomposition methods, rarely documented in academia but widely used in industry, offer fast and highly interpretable decompositions of metric changes, yet still fall short for complex metric systems. Additionally, because they are grounded in analytics and metric definitions rather than explicit modeling, they typically ignore causal dependencies.
In this work, we drew on these two representative lines of work (causal distribution-change methods and industry-style metric change decomposition)
and developed an attribution method that is interpretable to platform operators, scalable to large marketplace metric systems (130{+} metrics in our deployment), and grounded in a principled causal framework.
In summary, we make the following contributions:
\begin{itemize}
     \item We formalize an industry-style metric change decomposition method and extend it into a recursive metric-tree framework for multi-level root cause analysis in complex marketplace metric systems (Sec.~\ref{sec:tree_decomp}).
    \item We provide a theory- and simulation-based comparison with the GCM-based distribution-change method that clarifies their trade-offs in interpretability, scalability, causal validity, and attribution target misalignment (Sec.~\ref{sec:compare}).
    \item Guided by these insights, we propose a unified, causally informed attribution approach that injects structural causal information into the metric-tree framework and introduces an unbiased estimator for mean-change attributions, improving upon GCM-based estimates while preserving interpretability and fast computation (Sec.~\ref{sec:ProposedMethod}). We support this with mathematical proofs and validation on simulations in Sec.~\ref{sec:simu}, and a real-world marketplace application is provided in Sec.~\ref{sec:real-world}.
\end{itemize}

 \section{Background: Graphical Causal Model-Based Distribution Change Method (GCM-DC)}
To identify root causes, \cite{GCM2021} introduced a GCM-based approach by framing the root causing task as ``which mechanisms are responsible for the change in the marginal distribution of a target variable''. 
Given a correctly assumed causal graph $G$, a.k.a. causal DAG (causal Directed Acyclic Graph, \cite{neal2020causalInf}), over variables $X_1,\dots,X_m$, the joint distribution factorizes into causal mechanisms (conditional distributions) as
\[
P_X(x_1,\dots,x_m) \;=\; \prod_{j=1}^m P_{X_j \mid \mathrm{PA}_j}(x_j\mid \mathrm{pa}_j),
\]
where $\mathrm{PA}_j$ are the parents of $X_j$. A mechanism change means replacing some conditionals $P_{X_j \mid \mathrm{PA}_j}$ with new ones $\tilde P_{X_j \mid \mathrm{PA}_j}$, yielding a new joint distribution
\[
P_X^{T}(x) \;=\; \prod_{j\in T} \tilde P_{X_j \mid \mathrm{PA}_j}(x_j \mid \mathrm{pa}_j) 
\prod_{j\notin T} P_{X_j \mid \mathrm{PA}_j}(x_j \mid \mathrm{pa}_j),
\]
for a change set $T \subseteq \{1,\dots,m\}$. The new marginal of a target $X_k$ is then obtained by summing all variables but $x_k$: 
\[
\tilde P_{X_k}(x_k) \equiv P^{T}_{X_k}(x_k) \;=\; \sum_{\mathbf{x}_{\setminus k}}  P_X^{T}(x_1,\dots,x_n),
\]
so any change in the marginal $P_{X_k}$ to $\tilde P_{X_k}$ is explained by the subset of mechanisms that changed. 
Since attribution depends on the order in which mechanisms are changed, the method uses Shapley values to attribute the marginal change to individual mechanisms by averaging over all possible orderings of the change set $T$.




For root causing changes in the mean of the target variable $X_k$, let $\mathbb{E}_{X_k \sim P_{X_k}}[X_k]$ denote the mean of the variable $X_k$,  writing the mean change as $\mathbb{E}_{\tilde P_{X_k}}[X_k] - \mathbb{E}_{P_{X_k}}[X_k]$. 
The Shapley value contribution of a node $X_j$ to the mean change of target $X_k$ is
\begin{equation}
\label{eq:shapleyContr}
    \phi_j(\e)
\;=\; \sum_{T \subseteq M\setminus\{j\}} \frac{1}{m \binom{m-1}{|T|}} \, C_{\e}(j \mid T), \end{equation}
where $M$ denotes a set of m variables, and the marginal contribution $C_{\e}(j \mid T)$ of $X_j$ given a change set $T$ is defined as
\begin{equation}
\label{eq:marginalContr}
    C_{\e}(j \mid T)
    \;=\; \mathbb{E}_{X_k \sim P^{T\cup\{j\}}_{X_k}}[X_k]
        - \mathbb{E}_{X_k \sim P^{T}_{X_k}}[X_k]. 
\end{equation}
This method interprets the change in the target's mean as arising from changes in the local causal mechanisms and uses Shapley values to allocate the mean shift to individual variables in the causal graph in a principled, order-invariant way. Based on the ordering and magnitudes of the Shapley value contribution of each variable, root causes are identified.

\section{Metric-Tree Change Decomposition (MTCD)}
\label{sec:tree_decomp}
Metric decomposition is widely used across industries to explain metric changes, but existing treatments appear mostly in technical blogs~\cite{shao2023medium,halford2023kpi,hargrave2026dupont,microsoft2025KPItree} and have two key limitations. First, metric and attribution types are not organized into a systematic framework. Second, existing practical treatments either operate only at a single parent--child level or cover only the simplest attribution type in a simple tree setup. These limitations make it difficult for practitioners to derive a general algorithm that propagates attributions across common metric types in a multi-level tree while preserving exact additivity to the root.



We address these limitations. We first distinguish \emph{non-rate} metrics $y$ (e.g., revenue, page views) from \emph{rate} metrics $\bar{y}$ (e.g., average unit price (AUP), conversion rate(CVR)). For two periods, $t=0$ and $t=1$, let $z_0$, $z_1$, and $\Delta z := z_1-z_0$ denote the baseline value, new-period value, and change of any scalar quantity $z$.

\begin{figure}[ht]
    \centering
    \includegraphics[width=0.5\linewidth]{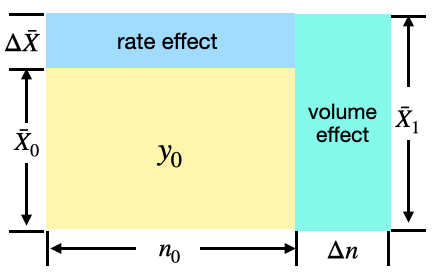} 
    \caption{Type 2 metric allocation strategy}
    \label{fig:decomp1}
\end{figure}

A \emph{metric decomposition} writes a target metric as a deterministic function of its components,
\[
y=f(\text{components}) \qquad \text{or} \qquad \bar{y}=f(\text{components}),
\]
typically through sums, products, or weighted averages defined by business logic. Its corresponding \emph{change decomposition} expresses the target metric change as
\[
\Delta y=\sum_j \text{contrib}_j \qquad \text{or} \qquad \Delta \bar{y}=\sum_j \text{contrib}_j,
\]
where the component contributions are interpretable and sum exactly to the observed change. Table~\ref{tab:CtC_types} summarizes the main metric decompositions and their associated change decompositions. In Type~2 (Fig.~\ref{fig:decomp1}), the allocation attributes the volume effect using the new-period price rather than the old-period price, reflecting a natural business intuition; in Type 4, $(\bar{y}_{k,0}-\bar{y}_0)$ centers subgroup rates around the baseline overall rate so that the share effect reflects whether higher- or lower-rate groups gain share \cite{shao2023medium}.

\begin{table*}[t]
\centering
\caption{Basic metric decompositions and their change decompositions.}
\label{tab:CtC_types}
\small
\begin{tabular}{p{0.23\textwidth}p{0.27\linewidth}p{0.43\linewidth}} 
\toprule
\textbf{Category} & \textbf{Metric decomposition} & \textbf{Change decomposition} \\
\midrule
Type 1: Non-rate(or rate), \newline 
(sum)
& $y = \sum_{k=1}^K y_k$ \newline
components are $y_1,\ldots, y_K$
& $\Delta y = \sum_{k=1}^K\Delta y_k,$ where
$\Delta y_k = y_{k,1} - y_{k,0}$. \newline
Effect of changes of $y_k$ is given by $\Delta y_k$. \\[0.35em]
\midrule
Type 2: Non-rate, \newline (product)
& $y = n \cdot \bar{X}$ \newline
components are $n$, $\bar{X}$ \newline
e.g., revenue = units sold$ \cdot$AUP
& $\Delta y = \mathrm{vol} + \mathrm{rate}$ (Fig. \ref{fig:decomp1}), \newline
 vol (volume effect of  $n$) = $\Delta n \cdot \bar{X}_1$, \newline
 rate (rate effect of $\bar X)= \Delta \bar{X} \cdot n_0.$ \\[0.55em]
\midrule
Type 3:Non-rate,  \newline (sum of products)
& $y = \displaystyle\sum_{k=1}^K n_k \cdot \bar{X}_k$ \newline
components are $n_k$, $\bar{X}_k$
& $\displaystyle\Delta y = \sum_{k=1}^K (\mathrm{vol}_k + \mathrm{rate}_k),$ \newline
$\mathrm{vol}_k  (\text{volume effect of } n_k) = \Delta n_k \cdot \bar{X}_{k,1},$ \newline $\mathrm{rate}_k (\text{rate effect of } \bar X_k) = \Delta \bar{X}_k \cdot n_{k,0}.$ \\
\midrule
Type 4: Rate,  \newline (weighted average)
& $\displaystyle\bar{y} = \sum_{k=1}^K p_k \cdot \bar{y}_k,$ \newline
where $\sum_{k=1}^K p_k =1$\newline
components are $p_k$, $\bar{y}_k$
& $\displaystyle\Delta \bar{y} = \sum_{k=1}^K (\mathrm{share}_k + \mathrm{rate}_k),$ \newline
$\mathrm{share}_k (\text{share effect of } p_k ) = \Delta p_k \cdot (\bar{y}_{k,0} - \bar{y}_0),$ \newline
$\mathrm{rate}_k (\text{rate effect of } \bar y_k) = \Delta \bar{y}_k \cdot p_{k,1}.$ \\
\bottomrule
\end{tabular}
\end{table*}

These four patterns define the local rules of our metric-tree change
decomposition (MTCD) framework. We first note that the parent--child direction and root/leaf terminology in a metric tree is reversed relative to the GCM formulation. Let $\mathcal{T}$ be a 
metric tree with root $r$, where each node $v\in\mathcal{T}$ denotes
a metric with baseline and new-period values $y_{v,0}$ and
$y_{v,1}$, and $\Delta y_v:=y_{v,1}-y_{v,0}$. Each child $c\in \mathrm{Ch}(v)$
has a decomposition type
${\tt type}(c)\in\{\text{Type1},\text{Type2},\text{Type3},\text{Type4}\}$ linking to its parent $v$ from Table~\ref{tab:CtC_types} and has a dimension label $\dim(c)$.  
We construct the metric tree under the MECE (mutually exclusive, collectively exhaustive) principle~\cite{tung2024MECE1,lee2018MECE2} to systematically link metrics and decompose target changes. 

The root contribution is initialized as $C_r=\Delta y_r$, and we can recursively apply the corresponding local change decomposition to each internal node from Table~\ref{tab:CtC_types} by propagating
\[
    C_c =
    \begin{cases}
    \mathrm{contrib_c} \cdot \frac{C_v}{\Delta y_v}, & \Delta y_v\neq 0,\\
    0, & \Delta y_v=0,
    \end{cases}
    \qquad c\in\mathrm{Ch}(v) \ \text{for all dim}(c) .
\]
This recursion preserves exact additivity within each decomposition
dimension and allocates the target change to all descendant nodes.
Non-root nodes with the largest absolute contributions are reported
as root causes. Appendix~\ref{Apex:MTCD_algo} provides the pseudo-code.

\begin{figure}[ht]
    \centering
    \includegraphics[width=\linewidth]{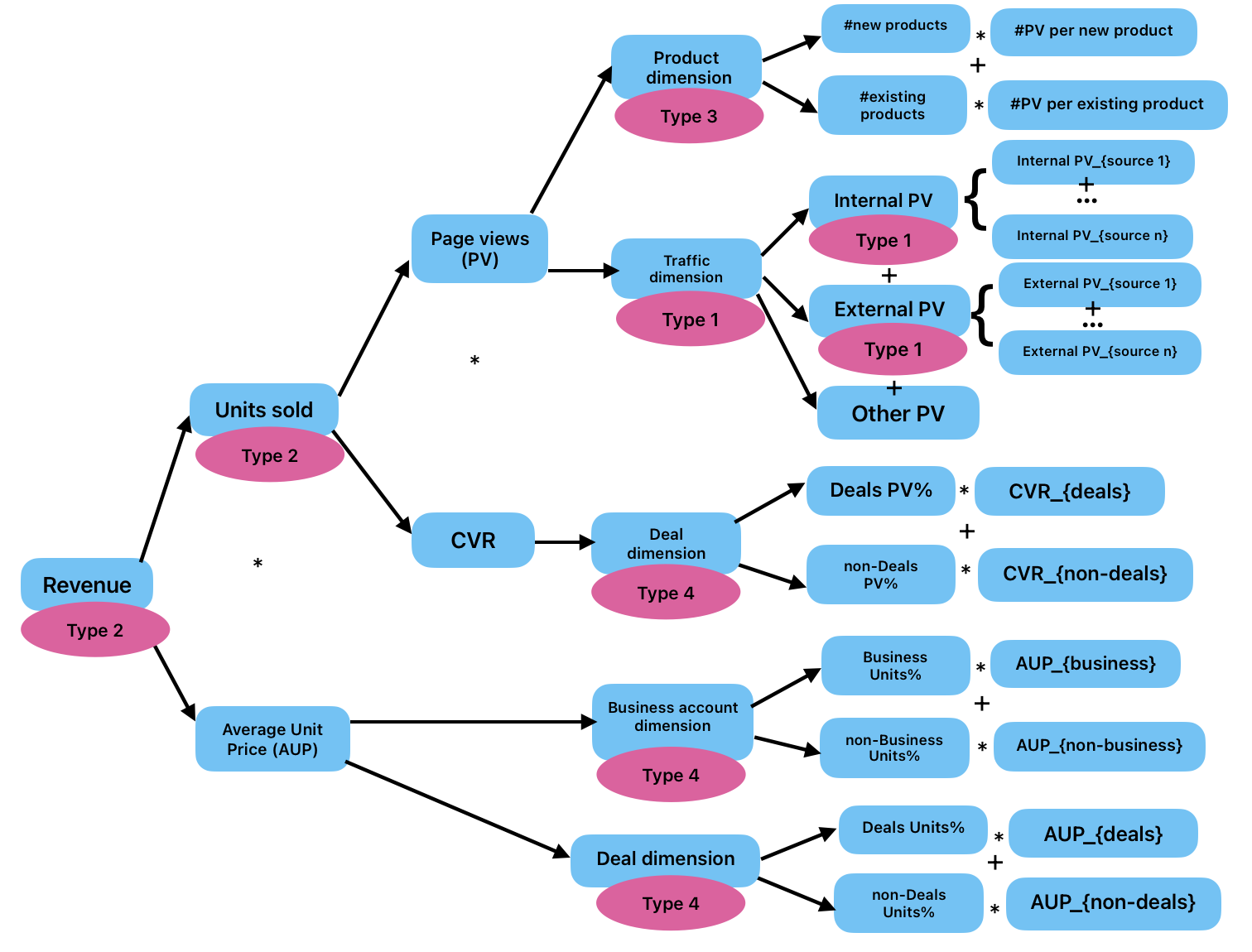}
    \caption{Complex metric tree example}
    \label{fig:bigTree}
\end{figure}

For illustration, revenue can be decomposed into units sold ($n$)
and average unit price (AUP) by Type~2, while AUP can be further
decomposed by dimensions such as deal status through Type~4. Using
this construction, we built a metric-tree system covering over 130
metrics, a subset of which is in Fig.~\ref{fig:bigTree}.




\section{Problem Statement}
\label{sec:limitations}
In a complex metrics system, both the GCM-DC method and the MTCD method offer clear benefits but also important limitations for interpretable and reliable root cause attribution. In particular, MTCD is simple to construct, highly interpretable, and essentially instantaneous to compute at scale. However, it has structural limitations:
\begin{enumerate}
    \item \textbf{Cross-branch dependencies not modeled.}
    The tree decomposition structure does not allow dependence among siblings or nodes of the same layer, hence not able to incorporate all causal relations: e.g., AUP may have impact on units sold. 
    \item \textbf{Incomplete coverage of non-decomposable metrics.}
    The framework cannot cover metrics that are not decomposable. Although much of the time, we can define metrics following the MECE rule and decompose consequently, but there are exceptions.
\end{enumerate}

The GCM-DC approach can fix both restrictions, but it also comes with some limitations of its own compared to MTCD:

\vspace{0.5em}
\noindent\textit{Known limitations of GCM modeling:}
\begin{enumerate}
    \item \textbf{Lower interpretability and lack of exact additivity in implementation.}
    Attributions are produced by a collection of conditional models and numerical procedures, making the method effectively a black box. Contributions typically do not sum exactly to the observed change in the target metric due to residual noise in each conditional model;
    \item \textbf{Dependence on the causal graph.}
    Validity hinges on the correctness of the assumed causal graph. Mis-specified edges, directions, or unmeasured confounders can make the inference unreliable, and the number of possible graphs grows exponentially (up to $2^{m(m-1)}$ edge configurations for $m$ nodes), making misspecification hard to avoid in practice, even with falsification tests \cite{eulig2025falsification}.
\end{enumerate}

\vspace{0.5em}
\noindent\textit{Limitations discovered in our later analyses and simulations:}
\begin{enumerate}
    \setcounter{enumi}{2}
    \item \textbf{Computational cost.}
    For realistic graphs (e.g., a 34-node system with daily data for 1 year), computing a point estimate using GCM-DC can take minutes, and confidence intervals via resampling can take over ten minutes, making the real-time root causing impossible. For comparison, MTCD is on the order of a second;   
    \item \textbf{Sample size requirements.}
    Reliable estimation requires substantial data as each conditional distribution needs to be estimated. Month-over-month comparisons with roughly 30 daily observations often yield high-variance and unstable attributions;
    \item \textbf{Estimand mismatch under unique causal ordering.} Even with the correct graph, GCM-DC's Shapley-based attributions diverge from the true contribution target when a unique causal ordering exists, because Shapley allocation averages over impossible orders instead;
    \item \textbf{Sensitivity to outliers.}    
    The use of residual resampling to approximate marginal distributions \cite{dowhyGCM2024} makes estimates sensitive to heavy-tailed residuals and outliers, which can further destabilize the results;
    \item \textbf{Limited granularity of structural changes.}  
    Structural / Mechanism changes at a node are often summarized as a single term (e.g., overall CVR distribution is changed), whereas MTCD can expose finer-grained subgroup effects (e.g., which segment of CVR is changed).
\end{enumerate}

\section{Formal Comparisons of GCM-DC and Decomposition Approach}
\label{sec:compare}
\subsection{Real-world Applications}
\begin{figure}[ht]
    \centering
    \includegraphics[width=1\linewidth]{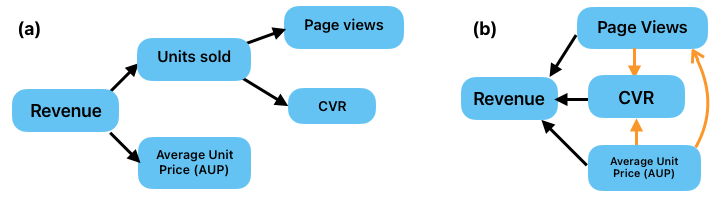}
    \caption{(a) Top-level decomposition tree; (b) Simple causal graph for revenue}
    \label{fig:4nodes}
\end{figure}

First, we illustrate the first two layers (Fig. \ref{fig:4nodes}(a)) for revenue change in a real use case. Based on domain knowledge, we define the causal graph in Fig. \ref{fig:4nodes}(b). We collected two years of daily data from a global e-commerce site and applied the GCM-DC method (using Fig. \ref{fig:4nodes}(b)) and MTCD (using Fig. \ref{fig:4nodes}(a)) to identify the root causes of year-over-year revenue changes for two vendors. Contribution values were rescaled to protect confidentiality. Because high variability made month-to-month GCM-DC results statistically insignificant, revealing its sample-size limitations, we restricted all real-world analyses to yearly comparisons.

\begin{figure}[ht]
    \centering
    \includegraphics[width=8.58cm]{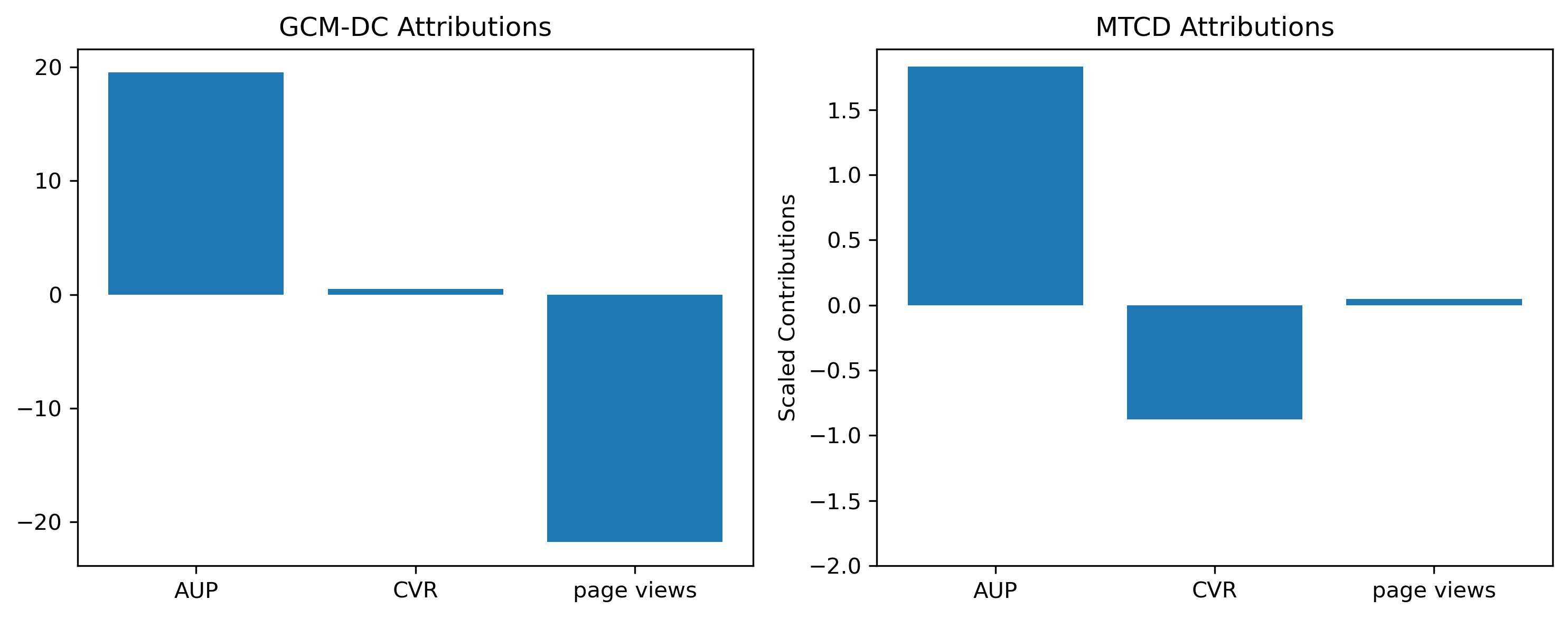}
    \caption{Vendor 1 contributions for year-over-year revenue change}
    \label{fig:vendor1}
\end{figure}

\begin{figure}[ht]
    \centering
    \includegraphics[width=1\linewidth]{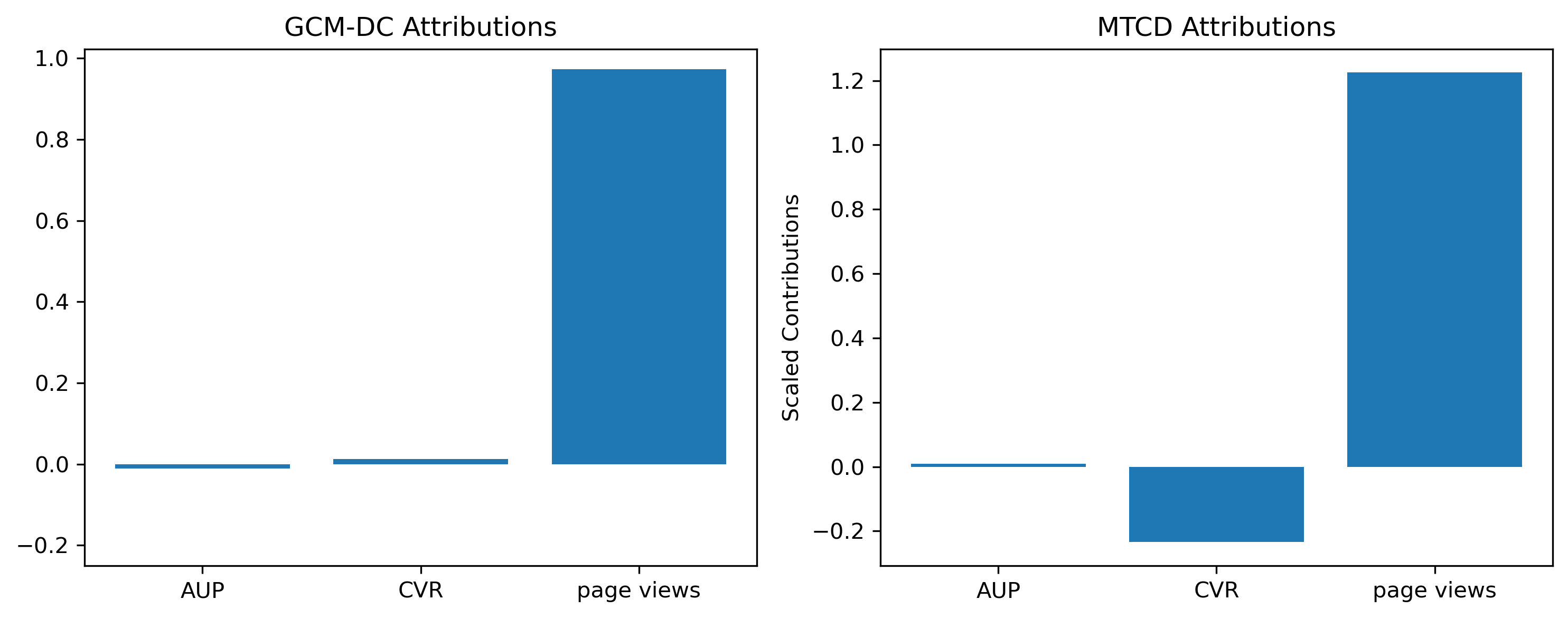}
    \caption{Vendor 2 contributions for year-over-year revenue change}
    \label{fig:vendor2}
\end{figure}

For the first vendor, we see from Fig. \ref{fig:vendor1} that contributions are swapped between CVR and page views under two approaches. For the second vendor (Fig. \ref{fig:vendor2}),  MTCD identified CVR as the second root cause, while GCM-DC did not. In summary, the results between GCM-DC and MTCD differ considerably, 
and it remains unclear which method is more trustworthy and why. This ambiguity necessitates further investigation of: (1) theoretical difference and linkage; and (2) simulations where the ground truth is known.

\subsection{Theoretical Proofs}
These two methods appear fundamentally different and stem from distinct methodological traditions, yet a detailed mathematical examination reveals their underlying connections and key differences. Without loss of generality, we will separate the findings into non-rate metrics and rate metrics. As for the non-rate metric, since Type 1 change decomposition is rather trivial, and Type 3 is a combination of Type 1 and 2, we will primarily focus on Type 2 decomposition.  

\subsubsection*{(I). Non-rate metric decomposition}
e.g., Revenue (Rev) = units sold (n) $\times$ AUP
    \subsubsection*{Differences:}
    \begin{enumerate}[label=(\roman*)]
        \item \label{item:scaleDiff} Different target scale (mean vs total change). Taking the same revenue example, the GCM-DC approach studies the mean change in revenue, while the decomposition-based approach allocates ``overall revenue change'' rather than the ``mean change''. However, either results can be adjusted to be on the same scale by {GCM-DC's contribution} $\times$ {sample size} or {Decomposition's contribution} / {sample size};
    
        \item GCM-DC uses Shapley allocation: calculate each node's average impact across all possible combinations of other nodes' change status, showing how much it contributes to the target metric change. 
        
        For example, contribution of units sold ($n$), using Eq. \eqref{eq:shapleyContr}-\eqref{eq:marginalContr}, 
        \begin{align}
        \label{eq:Shapley}
        & = \frac{1}{2} \cdot \text{impact of n on revenue given AUP not changed}  \nonumber \\
        & + \frac{1}{2} \cdot \text{impact of n on revenue given AUP changed} \nonumber \\
        & =     \frac{1}{2} \left( \e_{\mathrm{Rev}}^{\phi \cup n}[\mathrm{Rev}] - \e_{\mathrm{Rev}}^{\phi}[\mathrm{Rev}] \right)
            +\frac{1}{2} \left( \e_{\mathrm{Rev}}^{\mathrm{AUP} \cup n}[\mathrm{Rev}] - \e_{\mathrm{Rev}}^{\mathrm{AUP}}[\mathrm{Rev}] \right)
        \end{align}
        
        vs. Decomposition uses ordered allocation: pre-specified order of change impact. In the revenue case, AUP is assumed to be changed first, units sold is changed consequently;
        
        \item GCM-DC can incorporate additional causal relations among causes (which are sibling nodes in the metric tree): e.g., AUP impacts units sold by modeling units sold with a function of AUP and noise. 
    
    \end{enumerate}

    \subsubsection*{Underlying Connection:}
    \begin{proposition}
    \label{prop:non-rate_connection}
     The scale adjusted GCM-DC and decomposition approach is equivalent in expectation when causes are independent and ordered allocation (instead of Shapley) is used. 
    \end{proposition}
    Appendix \ref{Apex:non-rateSpecialCase} provides the proof.

\subsubsection*{(II). Rate metric decomposition}  
\label{sec:theory_rate}
e.g., $\text{AUP} = \text{business units}\% \ast \text{AUP}_b + \text{non-business units}\% \ast \text{AUP}_{nb},$
where subscripts $b$ and $nb$ are abbreviated for the business and non-business subgroup. 

    \subsubsection*{Differences:}
    \begin{enumerate}[label=(\roman*)]
        \item Slightly different target scale. Unlike non-rate decomposition, here GCM-DC analyzes the change of mean AUP between two periods, which is not mathematically equivalent to overall AUP change between two periods;

        \item Shapley allocation (used by GCM-DC) differs from ordered allocation (used by Decomposition), similar to the discussion in the non-rate decomposition above. 
    \end{enumerate}

    \subsubsection*{Underlying Connection:}
    \begin{proposition}
    \label{prop:rate_connection}
        In a linear setting, GCM-DC $\approx$ decomposition for sum of share nodes when ordered allocation (instead of Shapley) is used, or the rate's distribution is not changed. 
    \end{proposition}
This is proved in Appendix \ref{Apex:rateSpecialCase} and simulation studies in the next section (detail in Appendix \ref{Apex:sce4a} Scenario 4a). In summary, the main differences between the two methods are (a) the potential dependence among causes (i.e., sibling nodes in the metric tree) and (b) Shapley allocation vs ordered allocation.

\subsection{Simulation Studies} 
\label{sec:Simu_Ext}
This section verifies through simulations whether either method
is able to select the correct root causes when ground truth is available for various scenarios, reveals pros and cons of each method summarized in the problem statement section, and confirms the findings from the theoretical proofs section above.

\subsubsection*{I. Non-rate  metric simulation}

From Appendix \ref{Apex:simu_noRate}, in summary, when there is dependence among causes, MTCD would fail as expected. Under the correct DAG, GCM-DC can successfully identify root causes in the right order even with weak causal relationships, though contribution magnitudes diverge from the true contribution (Appendix Figs. \ref{fig:sce2a} \& \ref{fig:sce2b}) under the known causal ordering, revealing the estimand misalignment problem in Sec.~\ref{sec:limitations}. GCM-DC, as a gold-standard causal root-causing approach, still relies on correct causal DAG assumption, and fails when the causal DAG is mis-specified (Appendix Figs. \ref{fig:sce3a_AUP}-\ref{fig:sce3a_CVR}) or edges are missed (Appendix Fig. \ref{fig:sce3b}), underscoring the importance of correct graph assumptions. 
Additionally, a small sample size (e.g., 30) will make GCM-DC unreliable even under the correct DAG (Appendix Fig. \ref{fig:sce2a_n30}).

\subsubsection*{II. Rate metric simulation}
From Appendix \ref{Apex:simu_Rate}, it confirmed the two differences and the underlying connection stated in the above theoretical section. Moreover, given that GCM-DC models all parent nodes together with their joint child node vs. MTCD which decomposes rate metric into multiple dimensions separately in parallel (e.g., AUP in Fig. \ref{fig:bigTree}), we tested the hypothesis whether MTCD inflates the contribution of each node compared to GCM-DC. Under the assumption that multiple nodes are independent (e.g., subscription is independent of business account, Appendix \ref{Apex:simu_Sce5} Scenario 5), we verified that MTCD, which computes contributions in separate dimensions, can still estimate the true contributions fairly well similar to GCM-DC, hence proving that MTCD is a valid approach. 

In our business case, it is usually reasonable to assume such independence among causes (i.e., sibling nodes in MTCD) under Type 4 metrics. Our focus is thus to primarily tackle the problem observed in non-rate metric, specially the Type 2 metric. However, when such independence assumption is violated, a similar strategy proposed in the next section can be applied to Type 4 decomposition.

\section{Unified Root Causing Approach}
 \label{sec:ProposedMethod}
Given the limitations of both approaches listed in Section \ref{sec:limitations} together with the findings of previous section, we propose a unified approach that addresses most of those challenges. It formalizes root cause attribution as causal distribution change and provides a clean bridge between metric decomposition and full causal inference. This can be directly applicable to industrial metrics monitoring systems. 

\begin{figure}[ht]
    \centering
    \includegraphics[width=1\linewidth]{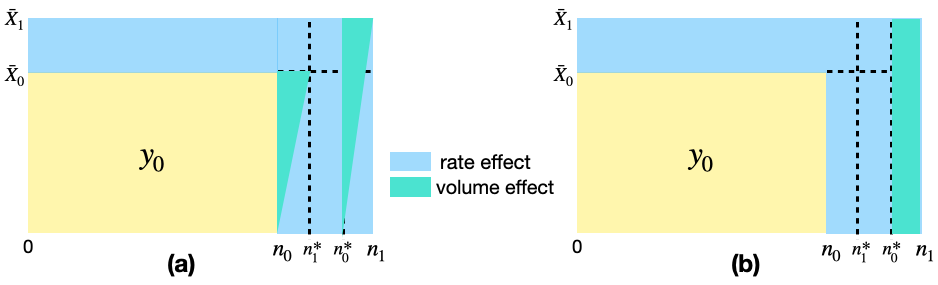}
    \caption{(a) Conceptual GCM-DC Allocation vs. (b) Proposed Allocation}
    \label{fig:GCM&unified allocation}
\end{figure}

Let us start with the original decomposition approach, non-rate metrics Type 2 uses the allocation in Fig. \ref{fig:decomp1}. The GCM-DC method incorporates both the causal relationship and the Shapley allocation. Based on the theoretical proofs section for the non-rate metric, we can conceptually represent its root causing mechanism of Eq. \eqref{eq:Shapley} geometrically in Fig. \ref{fig:GCM&unified allocation}(a),
where $n_1^* = \hat f_{1}(\bar{X}_0)$, $n_0^* = \hat f_{0}(\bar{X}_1)$
and $\hat f_0(\cdot)$ and $\hat f_1(\cdot)$ are to be learned from old-period and new-period data respectively. This is because in the case of AUP and units sold ($n$), the total contribution of $n$ comes from Eq. \eqref{eq:Shapley}:
\begin{align}
& =     \frac{1}{2} \left( \e_{\mathrm{Rev}}^{\phi \cup n}[\mathrm{Rev}] - \e_{\mathrm{Rev}}^{\phi}[\mathrm{Rev}] \right)
    +\frac{1}{2} \left( \e_{\mathrm{Rev}}^{\mathrm{AUP} \cup n}[\mathrm{Rev}] - \e_{\mathrm{Rev}}^{\mathrm{AUP}}[\mathrm{Rev}] \right) \nonumber\\
&\approx \frac{1}{2} \big[\frac{1}{J}\sum_{j=1}^J \big( n_{1,j}^* \cdot AUP_{0,j} - n_{0,j} \cdot AUP_{0,j}\big) \big] \nonumber\\
&+ \frac{1}{2} \big[\frac{1}{J}\sum_{j=1}^J \big( n_{1,j} \cdot AUP_{1,j} -  n_{0,j}^* \cdot AUP_{1,j} \big) \big],\label{eq:Shapley2}
\end{align}
where $j$ is the data granularity index, $j^{th}$ day, for example.
The two green triangles in Fig. \ref{fig:GCM&unified allocation}(a) correspond to two bracketed terms respectively in Eq. \eqref{eq:Shapley2} on a daily basis. 
Depending on the fitted two functions $\hat f(\cdot)$, $n_1^*$ and $n_0^*$ do not need to lie between $n_1$ and $n_0$. As a result, the estimated rate and volume effects can deviate from the original blue and green areas in Fig.~\ref{fig:decomp1}.

Inspired by GCM-DC, we build on MTCD by incorporating causal structure but with ordered allocation inferred by the causal path, akin to asymmetric Shapley methods~\cite{frye2020asymmetric,heskes2020causal}. Unlike prior work that restricts Shapley permutation weights for individual prediction explanations, our setting targets aggregate metric changes across two distributions and removes Shapley averaging entirely. When the added causal links have a unique causal order, a node's contribution is determined by its parents' status. Symmetric Shapley, by averaging over causally infeasible permutations, therefore targets a different attribution quantity than the ordered causal contribution, as confirmed in Sec.~\ref{sec:simu}. We instead follow the causal path and compute each node's contribution conditional on its causes under the new-period distribution, yielding a simpler attribution aligned with the intended causal estimand.

\begin{theorem}
\label{thm:unbias}
For three metrics $\{y, n, \bar X\}$ and $y=n \cdot \bar X$, under the correct local causal graph of $\bar X$ impacts $n$, and both $\bar X$ and $n$ impact $y$, if we let 
\begin{equation}
\label{eq:unbias}
     \hat {RC}(n|\bar X) = \frac{1}{J} \sum_{j=1}^J \bigl[ n_{1,j}-  \hat f_0(\bar X_{1,j}) \bigr]\cdot \bar X_{1,j}  ,    
\end{equation}
where the subscript $\{1,j\}$ denotes data from the new period, $\hat f_0(\cdot)$ is a function estimated from the baseline period between $n$ and $\bar X$, then $\hat {RC}(n|\bar X)$ is an unbiased estimator of the true root cause contribution of $n$ on the mean change of $y$. 

Remark: Keeping or removing the residual term within $\hat f_0(\bar X_{1,j})$ does not change the result under the assumption of the residual has mean of 0. 
\end{theorem}
The proof is provided in Appendix \ref{Apex:unbias}. In the running example, the contribution of $n$ to revenue can now be estimated by $\frac{1}{J}\sum_{j=1}^J \big(n_{1,j}-n_{0,j}^*\big)\cdot AUP_{1,j}$, where $n_{0,j}^*=\hat f_0(AUP_{1,j})$, same as the $2^{nd}$ bracket in Eq.~\eqref{eq:Shapley2}. Conceptually, this bracketed term corresponds to the green area in Fig.~\ref{fig:GCM&unified allocation}(b) at the daily level. Thus, the contribution of $n$ is its impact on revenue conditional on AUP’s new status, isolating the pure effect of $n$’s own mechanism change after removing AUP’s influence on both $n$ and revenue. The remainder blue area in Fig.~\ref{fig:GCM&unified allocation}(b) is attributed to AUP, through its direct effect $n_0\cdot \Delta \bar X$ and indirect effect $(n_0^*-n_0)\cdot \bar X_1$. Unlike GCM-DC, this approach does not require $n_1^*$ or Shapley averaging, yielding a simpler estimator that more accurately recovers root causes, as confirmed in the simulation section below.


Under a certain assumption, we can further simplify the calculation of the proposed estimator by 
\begin{proposition} 
\label{prop:simpT}
under the assumption when there is no relational change between $\bar X$ and $n$ in the new period (i.e., change is only from the intercept),
if we estimate 
\begin{equation}
\label{eq:simpT}
    \hat{RC}(n|\bar X) = \frac{1}J{}\big[n_1 - \sum_{j=1}^J \hat f_0(\bar X_{1,j}) \big] \cdot \bar X_1 ,
\end{equation}
where $n_1$ and $\bar X_1$ are the aggregated metrics at the period level. We can show that the simplified $\hat{RC}(n|\bar X)$ is also an unbiased estimator of the true root cause contribution of $n$ on the mean change of $y$.

\end{proposition}

The proof is provided in Appendix \ref{Apex:simpT}. Although aggregation may lead to an ill-defined estimator~\cite{aggregation2024}, we show that under the stated assumption, the simplified estimator in Eq.~\eqref{eq:simpT} remains unbiased and is computationally cheaper than Eq.~\eqref{eq:unbias}. Its form closely matches the original decomposition term $\Delta n\cdot \bar X_1$, with $n_0$ replaced by $\sum_{j=1}^J \hat f_0(\bar X_{1,j})$. This allows the remaining nodes be computed directly at the aggregate level, reducing complexity from $O(Jm)$ to $O(kJ+m)$ when $k$ causal dependence is added to an $m$-node tree. Regarding the proposition assumption, it is plausible when relationships among causes (i.e., sibling nodes in MTCD) are stable over limited time horizons or under relatively stable market conditions~\cite{budhathoki2022outlier}. In practice, it can be checked by fitting a pooled model, e.g.,$n=\beta_0+\beta_1\bar X+\beta_2 I(\text{period})+\beta_3 I(\text{period})\bar X+\epsilon$,
and testing $\beta_3=0$. If violated, we recommend the full estimator in Eq.~\eqref{eq:unbias}, which does not require this slope stability assumption. We note that both estimators target the mean change of $y$, and multiplying by the sample size $J$ yields the contribution to the total change.

The same strategy extends to other metric types. For Type 1 (sum), $\Delta y_k$ can be reassigned to its causes when dependencies exist. Type 3 (sum of products) combines Types 1 and 2, using the Type 2 rule within each product term and the Type 1 rule across groups. For Type 4 (weighted average), dependent share terms across dimensions (e.g., $\Delta p_{\text{ business}}$ and $\Delta p_{\text{deals}}$) can be reallocated along the assumed causal path, and if a rate affects its own share within a dimension, part of $\Delta p_k$ can likewise be reassigned to the rate contribution using a Type 2-style adjustment.

In general, Table \ref{tab:comparison} lists the comparison between the proposed approach and the GCM-DC approach. In order for a causal root causing while maintaining other properties, the proposed approach clearly outperforms the existing go-to causal approach. 

\begin{table}[ht]
\centering
\caption{Comparison between GCM-DC and Proposed Approach}
\label{tab:comparison}
\footnotesize
\begin{tabular}{
>{\raggedright\arraybackslash}p{2.2cm}
>{\raggedright\arraybackslash}p{2.5cm}
>{\raggedright\arraybackslash}p{2.5cm}
}
\toprule
\textbf{Evaluation Criteria} & \textbf{GCM-DC} & \textbf{Proposed Approach} \\
\midrule
Cross-branch dependence & \textcolor{green!60!black}{Yes} & \textcolor{green!60!black}{Yes} \\
\midrule
Non-decomposable metrics coverage & \textcolor{green!60!black}{Yes} & No (however, $>95\%$ are decomposable in practice) \\
\midrule
Interpretability & Low & \textcolor{green!60!black}{High} \\
\midrule
Causal graph correctness & entire DAG & \textcolor{green!60!black}{Only added edges} \\
\midrule
Computational cost & Relatively high (minutes for 30-40 nodes) & \textcolor{green!60!black}{Low ($\sim$ 1 second)} \\
\midrule
Sample size requirement & High (quarterly+) & \textcolor{green!60!black}{Lower (monthly is okay)} \\ 
\midrule
Attribution target alignment & Misaligned in some cases & \textcolor{green!60!black}{Aligned \& unbiased with true target attribution} \\
\midrule
Outlier sensitivity & Higher \newline (residual resampling) & \textcolor{green!60!black}{As usual} \\
\midrule
Granularity of structural changes & Not available \newline (e.g., $\text{CVR}_{\text{deals}}$)  & \textcolor{green!60!black}{Yes} \\
\bottomrule
\end{tabular}
\end{table}

\section{Simulations}
\label{sec:simu}
\begin{figure}[ht]
    \centering
    \includegraphics[width=0.4\linewidth]{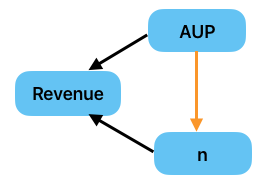}
    \caption{3 nodes causal DAG}
    \label{fig:3nodes}
\end{figure}

In this section, we use the same motivating example of AUP, units sold($n$) and revenue to simulate data, and allow potential interaction between AUP and $n$ using the causal graph as in Fig. \ref{fig:3nodes}. 
In the baseline period, we start by simulating AUP from a normal distribution, simulate $n$ using a function of AUP with a normal noise distribution, and let revenue $= n \cdot$ AUP. In the new period, we increase AUP and/or $n$ on different scales to test how the proposed unified approach performs compared to the original decomposition approach, the GCM-DC approach, and the true root causes attributions. Simulation parameters are calibrated from real-world data to make the synthetic data realistic in scale, while retaining full control over the causal mechanism changes and ground-truth attributions. Both periods have a sample size of 100.

To compute the contribution using the proposed approach, we first need to learn the relationship $f_0(AUP)$ from the simulated data in the baseline period, then instead of using the original decomposition formula of $\Delta n \cdot AUP_1$, we will use $(n_1 - n_0^*) \cdot AUP_1$ where $n_0^*$ can be estimated by $\sum f_0(AUP_{1j})$. This corresponds to the simplified estimator in Eq. \eqref{eq:simpT} before scaling.  We repeat the simulations 100 times, calculate the average across 100 runs, and compute 95\% confidence intervals for each method. As stated earlier, for both the original decomposition and the proposed approach, we also scaled the results by sample size to make them comparable with the GCM-DC results. 

\begin{table}[ht]
    \centering
    \caption{Simulation Cases}
        \begin{tabular}{l c c}
        \toprule
         & \shortstack{Causal relationship \\ for $n$\&AUP} & True root causes \\
        \midrule
        case 1a (Fig. \ref{fig:hybrid1a}) & \multirow{2}{*}{linear}    & AUP  \\
        case 1b (Fig. \ref{fig:hybrid1b}) &                            & AUP \& $n$  \\
        \midrule
        case 2a (Fig. \ref{fig:hybrid2a}) & \multirow{2}{*}{quadratic} & AUP  \\
        case 2b (Fig. \ref{fig:hybrid2b}) &                            & AUP \& $n$ \\
        \bottomrule
        \end{tabular}
\label{tab:simu_case}
\end{table}

\begin{figure}[ht]
    \centering
    \includegraphics[width=1\linewidth]{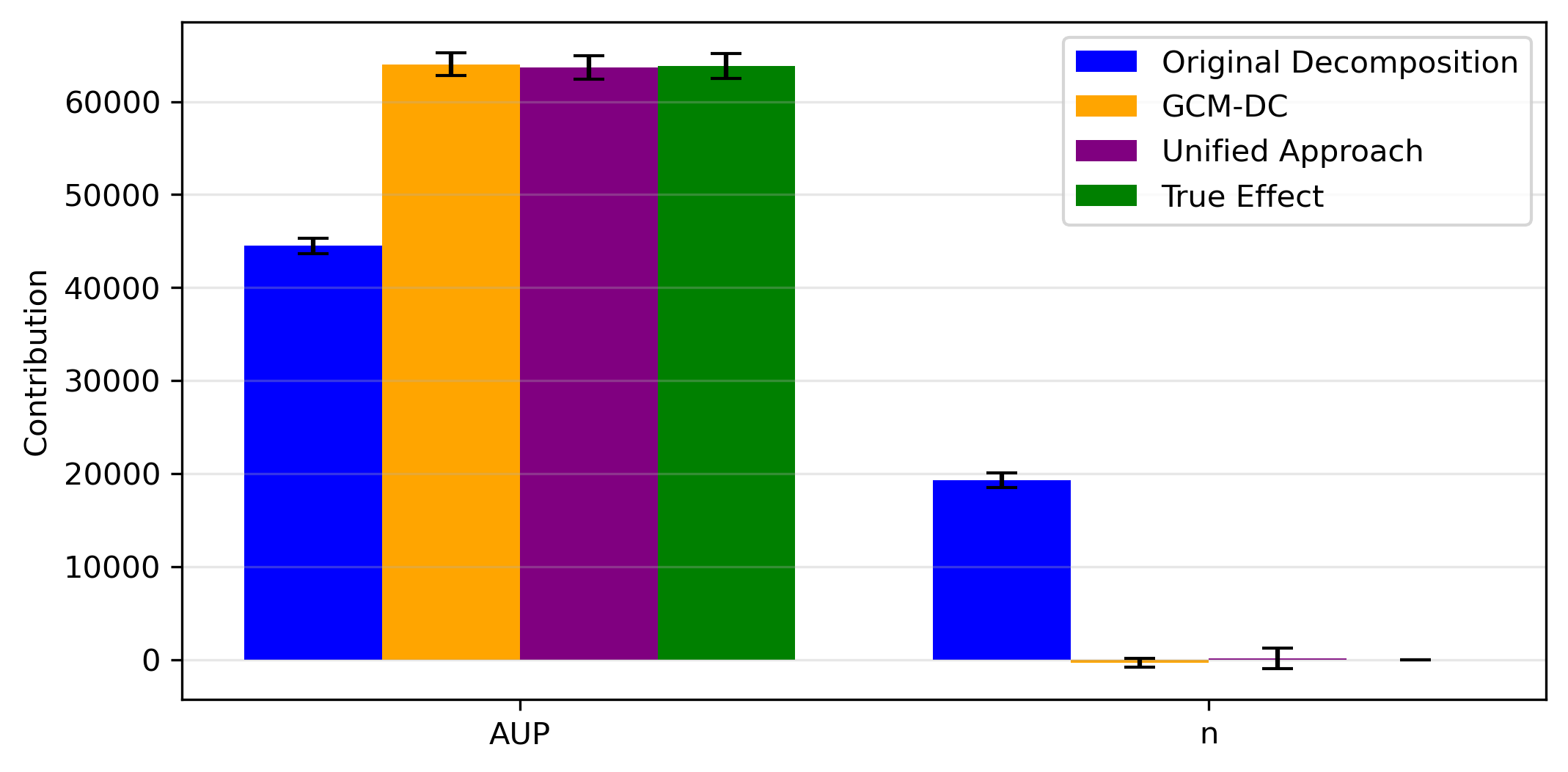}
    \caption{Case 1a. Linear, only AUP changed}
    \label{fig:hybrid1a}
\end{figure}
\begin{figure}[ht]
    \centering
    \includegraphics[width=1\linewidth]{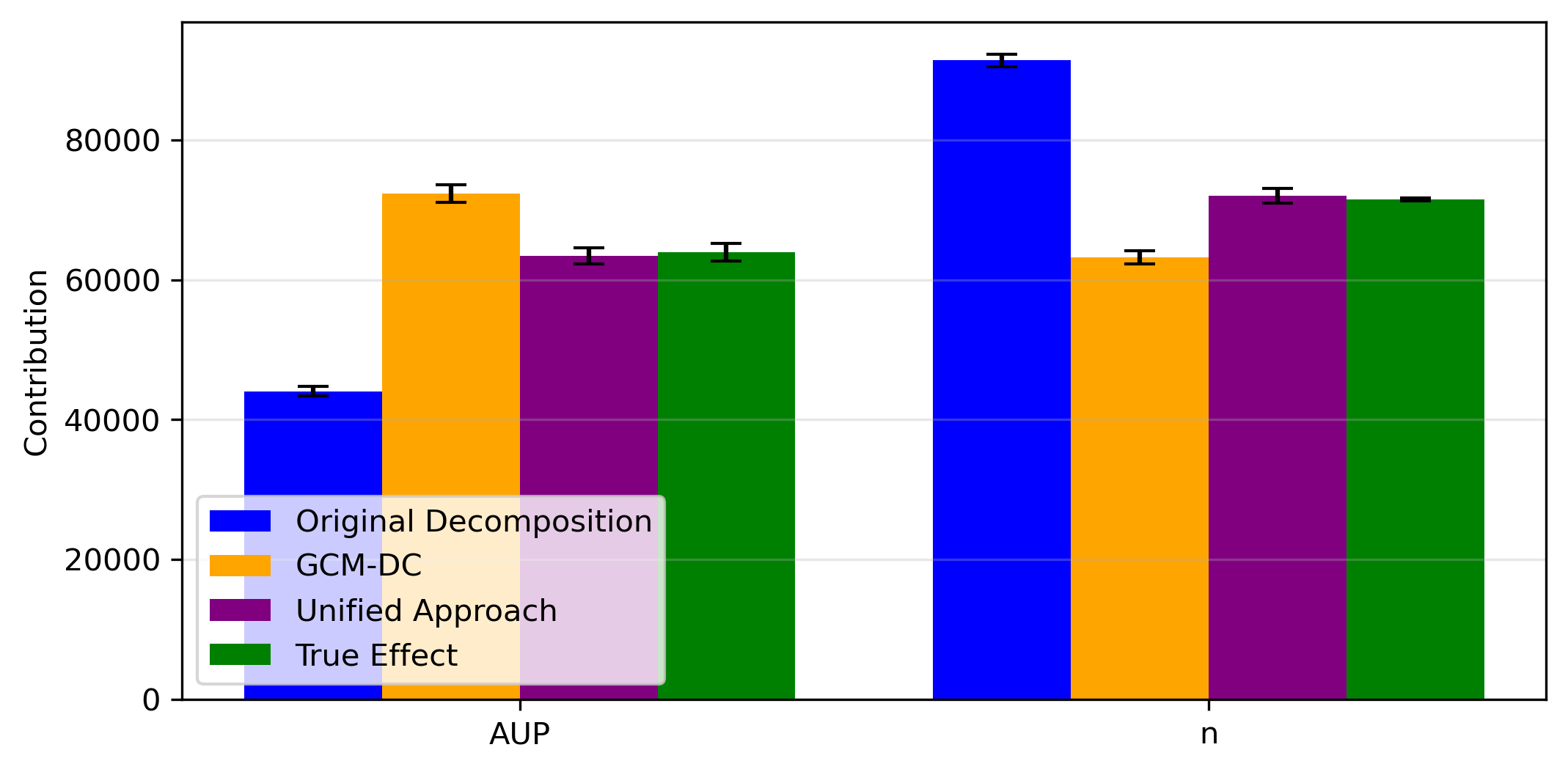}
    \caption{Case 1b. Linear, both AUP and n changed}
    \label{fig:hybrid1b}
\end{figure}
\begin{figure}[ht]
    \centering
    \includegraphics[width=1\linewidth]{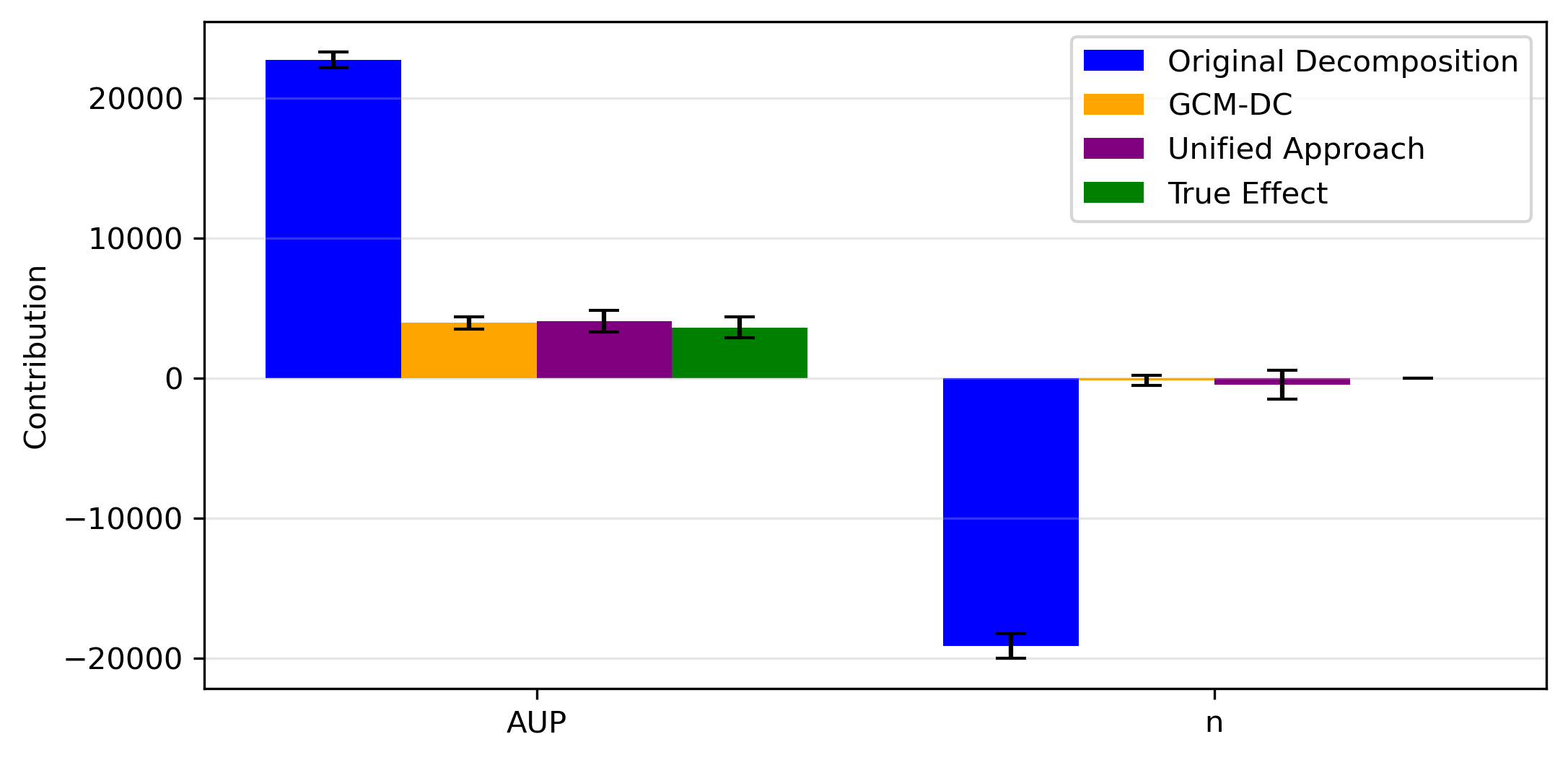}
    \caption{Case 2a. Quadratic, only AUP changed}
    \label{fig:hybrid2a}
\end{figure}
\begin{figure}[ht]
    \centering
    \includegraphics[width=1\linewidth]{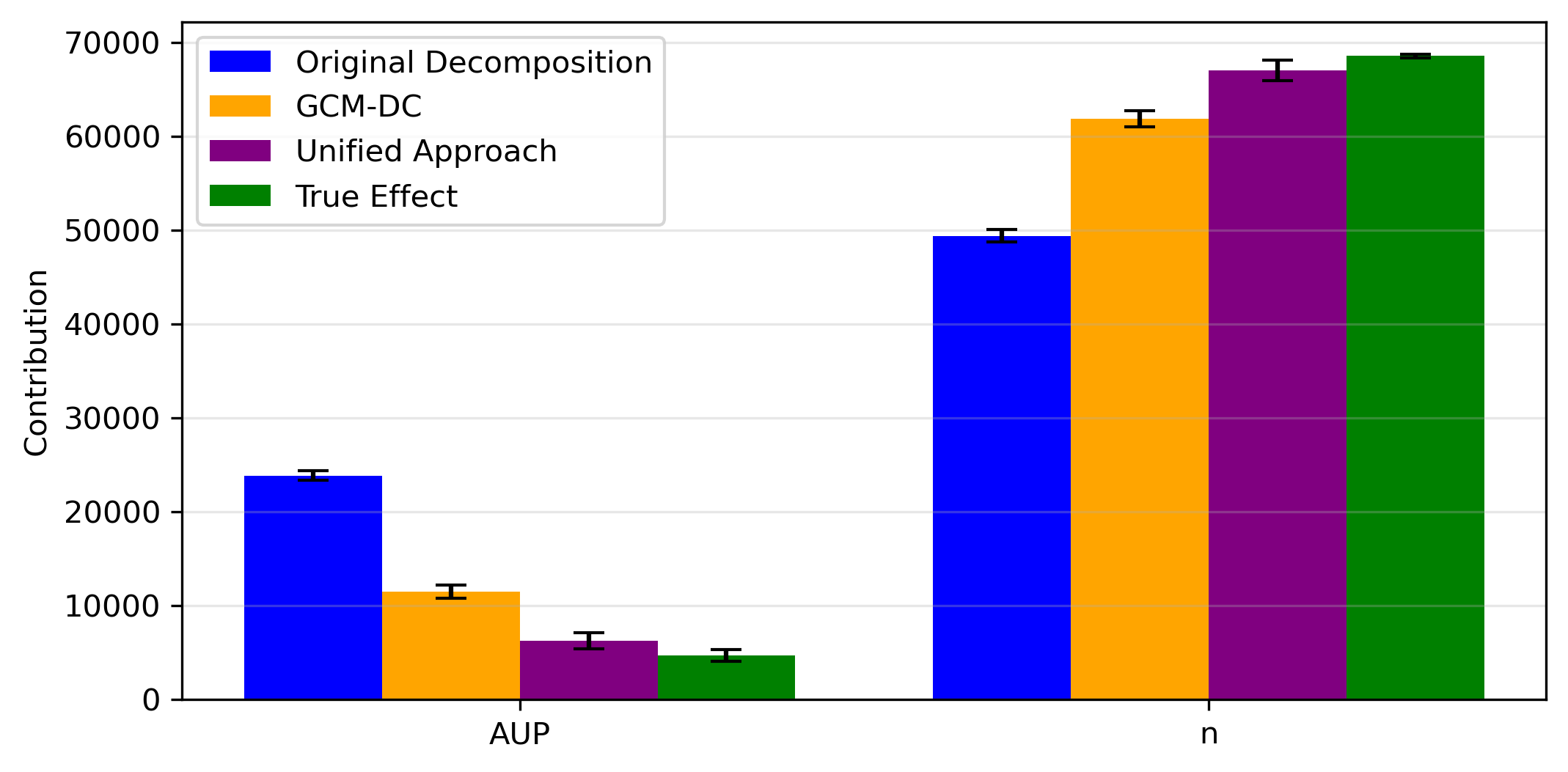}
    \caption{Case 2b. Quadratic, both AUP and n changed }
    \label{fig:hybrid2b}
\end{figure}

We performed simulations in 4 different cases (Table \ref{tab:simu_case}). From the results of Figs. \ref{fig:hybrid1a} and \ref{fig:hybrid2a}, the original decomposition incorrectly identified $n$ as the root cause because it did not take into account the causal dependence between AUP and $n$. Both the unified approach and the GCM-DC are very similar to the truth. This is expected when only one factor changes, Shapley allocation equals ordered allocation. 
From the results of Figs. \ref{fig:hybrid1b} and \ref{fig:hybrid2b} where both variables change, this is where the Shapley value may suffer from reduced accuracy, as it does not utilize the information of the causal path. In these two cases, the proposed unified approach is better than the GCM-DC approach and is closest to the true root causes. Again, if the original decomposition approach is used, the root causes are far from the truth, or in the wrong order. We also used Eq. \eqref{eq:unbias} instead of its simplified form to estimate the contributions and obtained nearly identical results, so we omit the plots. We purposely used a low-dimensional case to illustrate the essence of the problem and improvement, and because they are MTCD’s recursive building blocks, validity extends recursively to larger structures.

\section{Real-World Marketplace Application}
\label{sec:real-world}

\begin{figure}[ht]
    \centering
    \includegraphics[width=0.8\linewidth]{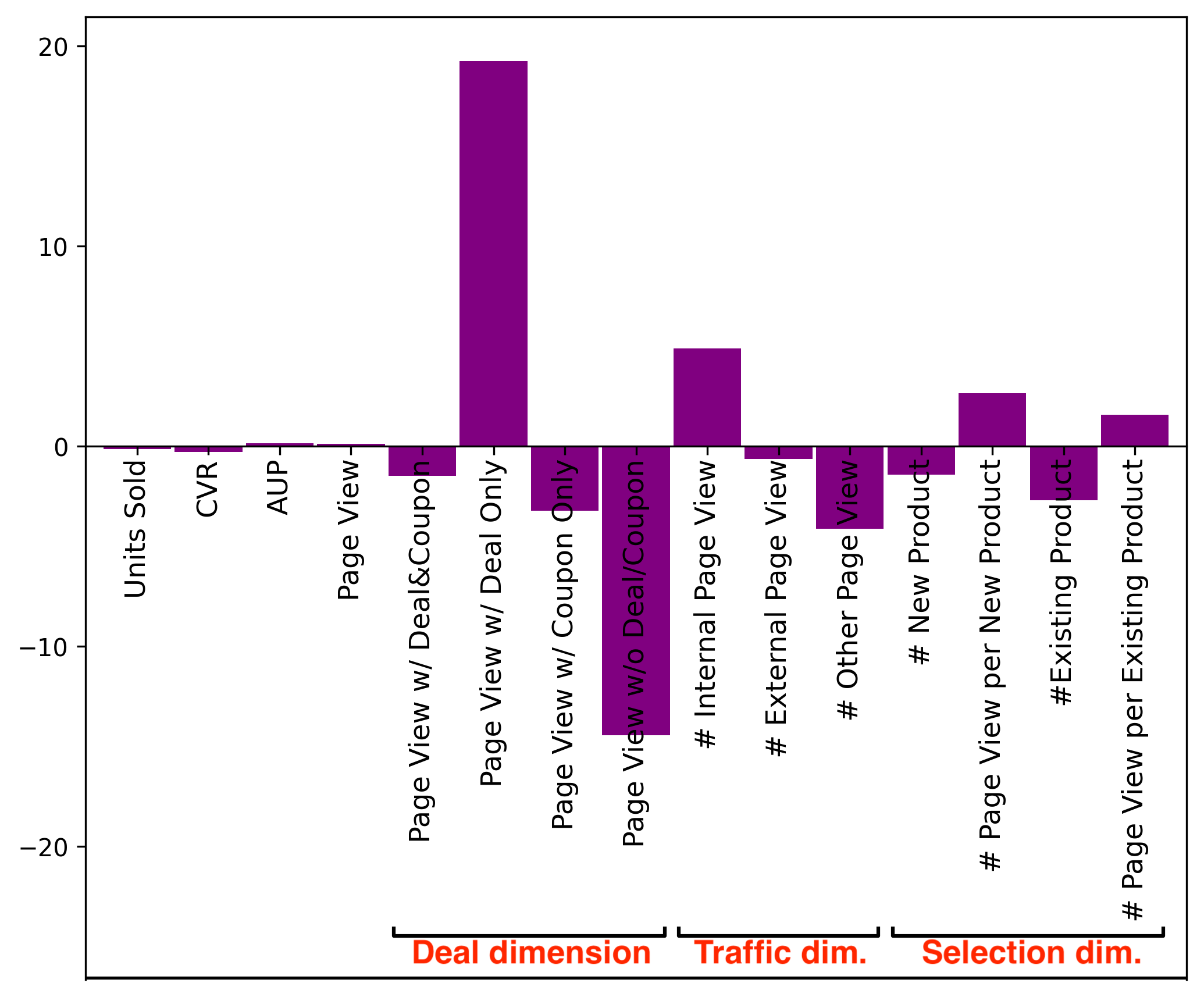}
    \caption{Real-world application}
    \label{fig:hybrid_realcase_bigTree}
\end{figure}

Our proposed method is currently deployed as a real-time root cause engine with flexible time and selection scopes, covering $>$130 metrics. As an illustration, we conduct a year-over-year comparison for a global e-commerce vendor using two years of daily data. We add only causal edges strongly supported by domain knowledge: AUP $\rightarrow$ units sold and page views $\rightarrow$ conversion rate. The runtime is 1.6 seconds on a standard laptop, approximately 400 times faster than GCM-DC under a comparable configuration.
In Fig. \ref{fig:hybrid_realcase_bigTree}, less important dimensions/leaf node drivers are omitted for clarity. We further rescaled by a constant to preserve data confidentiality. 
Applying the proposed approach, the top three positive drivers are page view (PV) with deal only, \#internal PV, and \#PV per new product, while the top three negative drivers are PV without deal or coupon, \#other PV, and PV with coupon only, with deal, traffic, and selection as the three main driver segments.
This more accurate attribution helps vendors/sellers quantify contribution sizes and prioritize growth actions.


\section{Concluding Remarks}



In this paper, we proposed a unified method for root cause analysis in complex metric systems that is causally informed, business interpretable, computationally efficient, and unbiased when the relevant causal ordering is unique. The method preserves the operational advantages of MTCD while incorporating sparse local causal structure to correct cross-branch attribution errors and prevents the estimand misalignment in GCM-DC. In production, modeling and maintenance costs remain negligible because modeling is required only for the sparse set of added causal edges, while the rest of the tree remains a model-free MTCD. This is substantially lighter than GCM-DC, which fits conditional models across the entire DAG.

The approach, although much less restrictive than the GCM-DC approach, relies on correct assumptions for the added \textit{local} dependent relationships. In practice, we rely on domain knowledge, temporal ordering of metrics, and conditional falsification checks \cite{eulig2025falsification} to validate those added causal links. Because this approach is still subject to ``standard'' sensitivity to outliers, it suggests future directions including outlier handling methods for root causing, as well as extending the framework to handle temporal dependence in production monitoring systems.

\bibliographystyle{ACM-Reference-Format}
\bibliography{myBib}

\appendix
\section*{Appendices}
\section{Metric-Tree Change Decomposition (MTCD): Pseudo-code}
\label{Apex:MTCD_algo}

Algorithm~\ref{alg:mtcd-dim} implements the recursive MTCD
procedure introduced in Sec.~\ref{sec:tree_decomp}. We use the same
notation as in the main text. In the implementation, the call to $\textsc{BaseDecomp}$ would be computed once per dimension $k$ and reused for all children $c$ with $dim(c)=k$. 

\begin{algorithm}[ht]
\caption{Metric-Tree Change Decomposition (MTCD)}
\label{alg:mtcd-dim}
\begin{algorithmic}[1]
\Require Tree $\mathcal{T}$ with root $r$; node values $y_{v,0},y_{v,1}$;
        decomposition types ${\tt type}(c)$; dimension label $dim(c)$
\Ensure Contribution $C_v$ for each node $v \in \mathcal{T}$, such that $\sum_{u \in {\rm Ch}(v),\,\dim(u)=k} C_u = C_v$ for all dimensions within node $v$. 
\Procedure{Propagate}{$v$}
\Statex \Comment{A function taking a node $v$, calculating contribution of its children nodes $c$ and looping through the children nodes.}
    \If{$v$ is leaf}
        \State \Return
    \EndIf
    \State $\Delta y_v \gets y_{v,1} - y_{v,0}$
    \ForAll{$c \in {\rm Ch}(v)$}
        \State $k \gets \dim(c)$
        \State $\bigl(\text{contrib}_{c}\bigr)_{u \in {\rm Ch}(v),\,\dim(u)=k}$
        \Statex \hspace{2em} $\gets \textsc{BaseDecomp}\Bigl(
                    {\tt type}(c),\ (y_{u,0},y_{u,1})_{u \in \{v\}\cup{\rm Ch}(v),\,\dim(u)=k}
                \Bigr)$
        \Statex \Comment{Establish parent-children local decomp. per dimension}
            \If{$\Delta {y_v} \neq 0$} \Comment{Avoid dividing by zero}
            \State $C_c \gets \text{contrib}_{c} \cdot \dfrac{C_v}{\Delta {y_v}}$ 
        \Statex \Comment{Scale by the parent's importance}
        \Else
            \State $C_c \gets 0$
        \EndIf
        \State \Call{Propagate}{$c$}
    \EndFor
\EndProcedure
\State 
\State $\Delta y_r \gets y_{r,1} - y_{r,0}$, \quad $C_r \gets \Delta y_r$ 
\Statex \Comment{Define root's contribution as its $\Delta$}
\State \Call{Propagate}{$r$} \Comment Start by calling the function on $r$
\end{algorithmic}
\end{algorithm}

\section{Proofs}
\subsection{Proof of Proposition \ref{prop:non-rate_connection}}
\label{Apex:non-rateSpecialCase}
Using the same example of Revenue (Rev) = units sold (n) $\times$ AUP, assuming the causal graph in Fig. \ref{fig:3nodes}, if we use ordered allocation (AUP first and $n$ later) rather than Shapley value allocation, Eq.~\eqref{eq:Shapley} reduces to 
\begin{equation}
\label{eq:nonRate_ordered}
    \e_{\mathrm{Rev}}^{\mathrm{AUP} \cup n}[\mathrm{Rev}] - \e_{\mathrm{Rev}}^{\mathrm{AUP}}[\mathrm{Rev}].
\end{equation}
\eqref{eq:nonRate_ordered} can then be estimated by 
\begin{equation}
\label{eq:nonRateSPpecialCase_est}
    \frac{1}{J}\sum_{j=1}^J \big( n_{1,j} \cdot AUP_{1,j} -  n_{0,j}^* \cdot AUP_{1,j} \big) ,
\end{equation}
where $n_{0,j}^*$ is the units sold that would have changed due to the change of AUP for record $j$. 
In the special case where AUP is independent of units sold, all $n_{0,j}^* = n_{0,j}$, \eqref{eq:nonRateSPpecialCase_est} reduces to 
\begin{equation}
\label{eq:nonRateSpecialCase_est2}
    \frac{1}{J}\sum_{j=1}^J \big( n_{1,j} -  n_{0,j}\big) \cdot AUP_{1,j}.
\end{equation}
Taking the expectation of \eqref{eq:nonRateSpecialCase_est2} results in 
\begin{equation}
\label{eq:nonRateSpecialCase_est3}
    \frac{1}{J}\sum_{j=1}^J \e [\big( n_{1,j} -  n_{0,j}\big) \cdot AUP_{1,j}] \\
    = \frac{1}{J}\sum_{j=1}^J\e [ n_{1,j} -  n_{0,j}] \cdot \e[AUP_{1,j}],
\end{equation}
where the last equation holds due to the special assumption that AUP is independent of units sold. Likewise, taking the expectation of the scaled decomposition formula of $\Delta n \cdot AUP_1 / J$ gives
\begin{align}
\label{eq:nonRateSpecialCase_est4}
    \e \big[\frac{1}{J}\sum_{j=1}^J \big( n_{1,j} -  n_{0,j}\big) \cdot AUP_1 \big] 
    &= \frac{1}{J}\sum_{j=1}^J \e [ n_{1,j} -  n_{0,j}] \cdot \e[AUP_{1}] \nonumber\\
    &= \frac{1}{J}\sum_{j=1}^J \e [ n_{1,j} -  n_{0,j}] \cdot \e[AUP_{1,j}].
\end{align}
The second last equation holds due to the independence assumption, and the last equation holds because $\e[AUP_{1}] = \e[AUP_{1,j}] = \e\big[\frac{\sum_j \text{Revenue}_{1,j}}{\sum_j n_{1,j}}\big]$ (formal proof can be found in the proof in \ref{Apex:simpT} Equations \eqref{eq:simT_proof6} and \eqref{eq:simT_proof7}). This completes the proof.

\subsection{Proof of Proposition \ref{prop:rate_connection}}
\label{Apex:rateSpecialCase}
Let $p_b$ and $p_{nb}$ denote business units\% and non-business units\% respectively, and use the AUP example in Sec.~\ref{sec:theory_rate}(II): $\text{AUP} = p_b \cdot \text{AUP}_b + p_{nb} \cdot \text{AUP}_{nb}$. In the GCM realm, understanding the impact of the business account on overall AUP is equivalent to modeling $p_b$ directly with overall AUP. If we assume that the rate node AUP is linearly impacted by its cause $p_b$, to compute $p_b$'s impact on AUP, GCM-DC followed the Shapley allocation's strategy by first computing two scenarios:
(1) when nothing is changed:
\begin{equation}
\label{eq:AUP_T0}
    \e_{AUP}^{p_b}[AUP] - \e_{AUP}^{\emptyset}\!\left[AUP\right]
            = \left(\beta_0 + \beta_1 P_{b,1}\right) - \left(\beta_0 + \beta_1 P_{b,0}\right)
            = \beta_1 \Delta P_b;
\end{equation}
(2) when AUP's mechanism is changed:
\begin{align}
    \label{eq:AUP_T1}
   \e_{AUP}^{p_b \cup AUP}\left[AUP\right] - \e_{AUP}^{AUP}\left[AUP\right]
            &= \left(\beta_0^* + \beta_1^* P_{b,1}\right) - \left(\beta_0^* + \beta_1^* P_{b,0}\right) \nonumber\\
            &= \beta_1^* \Delta P_b.
\end{align}
Thus the Shapley value contribution of $p_b$ equals the average of \eqref{eq:AUP_T0} and \eqref{eq:AUP_T1}, which is
\begin{align} 
\label{eq:rate_GCM}
        &= \tfrac{1}{2}\,\Delta P_b\,(\beta_1 + \beta_1^*) \nonumber \\[4pt]
        &\approx \tfrac{1}{2}\,\Delta P_b\,
            \Big[
                \big(AUP_{b,0} - AUP_{nb,0}\big)
                + \big(AUP_{b,1} - AUP_{nb,1}\big))
            \Big],
\end{align}
where the last approximation in \eqref{eq:rate_GCM} holds true because
\begin{equation}
\begin{aligned}
     \mathrm{AUP} &= p_b \,\mathrm{AUP}_b + p_{nb} \,\mathrm{AUP}_{nb} \\
    &= \mathrm{AUP}_{nb} + p_b \left( \mathrm{AUP}_b - \mathrm{AUP}_{nb} \right),
\end{aligned}
\end{equation}
and when a linear regression is estimated using granular observations, $\beta_0 = \e[\mathrm{AUP}_{nb}]  $ and $\beta_1 = \e[\mathrm{AUP}_b - \mathrm{AUP}_{nb}]$ holds true (\cite{weisberg2005regression}). 

On the other hand, in the decomposition paradigm, 

(i) contribution of $p_b = \Delta p_b \left( AUP_{b,0} - AUP_0 \right)$,

(ii) contribution of $p_{nb} = \Delta p_{nb} \left( AUP_{nb,0} - AUP_0 \right)$,\\
because $\Delta p_{nb} = -\Delta p_b$, adding (i) and (ii) equals
\begin{equation}
\label{eq:rate_specialCase}
\begin{aligned}
    &= \Delta p_b \left( AUP_{b,0} - AUP_0 \right) - \Delta p_b \left( AUP_{nb,0} - AUP_0 \right)\\ 
    &= \Delta p_b \left( AUP_{b,0}  - AUP_{nb,0}  \right).
\end{aligned}
\end{equation}
When AUP's distribution is not changed (i.e., $\beta_1 = \beta_1^*$), \eqref{eq:rate_GCM} and \eqref{eq:rate_specialCase} are equivalent. When ordered allocation is used (i.e., use the first scenario in \eqref{eq:AUP_T0} instead of both scenarios), \eqref{eq:AUP_T0} $\approx$ \eqref{eq:rate_specialCase}. The proof is complete.

\subsection{Proof of Theorem \ref{thm:unbias}}
\label{Apex:unbias}
Given the rational stated in Section \ref{sec:ProposedMethod}, when a reasonable causal graph is available, or the sequence of metrics (``players'') is fixed, one variable's contribution on the target variable depends only on its parent node(s)' (i.e., causes) status. Hence, in the case of the Type 2 non-rate metric, given the provided causal graph, it is natural to define the true root cause as $\e_y^{\bar X \cup n}[y] - \e_y^{\bar X}[y]$. We are ready to prove the unbiasedness. Take the expectation of our proposed estimator,
\begin{align}
\label{eq:unbiasProof1}
     \e[\hat {RC}(n|\bar X)] 
    &= \e\big[\frac{1}{J} \sum_{j=1}^J \big( n_{1,j} \cdot \bar X_{1,j}-  \hat f_0(\bar X_{1,j}) \cdot \bar X_{1,j} \big) \big] \nonumber \\
    &= \e\big[ n_{1,j} \cdot \bar X_{1,j} \big]-  \e\big[\hat f_0(\bar X_{1,j}) \cdot \bar X_{1,j} \big].
\end{align}
The first term in \eqref{eq:unbiasProof1} is equivalent to $\e_y^{\bar X \cup n}[y]$ by definition as both $n$ and $\bar X$ are values under the new distribution, and $y=n\cdot \bar X$ holds true at any given point. Now, focusing on the second term, we start by working backward from the final term of $\e_y^{\bar X}[y]$, let us define $f_0(\bar X_{1,j},\epsilon_{0,j})$ as the true $n$ that would have changed under the new distribution of $\bar X$, 
so
\begin{align}
     \e_y^{\bar X}[y]
    &\equiv \e_y^{\bar X} \big[ f_0(\bar X_{1,j},\epsilon_{0,j}) \cdot \bar X_{1,j} \big] \nonumber \\
    &= \e_y^{\bar X} \big[ [f_0(\bar X_{1,j}) + \epsilon_{0,j}] \cdot \bar X_{1,j} \big] \label{eq:additiveNoise} \\ 
    &= \e_y^{\bar X} \big[ f_0(\bar X_{1,j})\cdot \bar X_{1,j} \big] + \e_y^{\bar X} [\epsilon_{0,j} \cdot \bar X_{1,j}] \nonumber \\
    &= \e_y^{\bar X} \big[ f_0(\bar X_{1,j})\cdot \bar X_{1,j} \big] + \e_y^{\bar X} [\epsilon_{0,j}] \cdot \e_y^{\bar X}[\bar X_{1,j}] \label{eq:IndpNoise} \\
    &= \e_y^{\bar X} \big[ f_0(\bar X_{1,j})\cdot \bar X_{1,j} \big]. \label{eq:unbiasProof2}
\end{align}
Eq.~\eqref{eq:additiveNoise} is true under the typical additive noise model, and \eqref{eq:IndpNoise} is true because each noise term $\epsilon$ is commonly assumed to be independent of $\bar X$ variable. And under the assumption of $\e[\epsilon_{0,j}] =0$, \eqref{eq:unbiasProof2} is true. Regarding the second term of \eqref{eq:unbiasProof1}, under the assumption of modeling estimation of $\e [\hat f_0(\bar X_{1,j}) |\bar X_{1,j}] = f_0(\bar X_{1,j})$, 
\begin{align}
\label{eq:unbiasProof3}
     \e\big[\hat f_0(\bar X_{1,j}) \cdot \bar X_{1,j} \big]
    &= \e\big[\e [\hat f_0(\bar X_{1,j}) \cdot \bar X_{1,j} |\bar X_{1,j}]\big] \nonumber \\
    &= \e \big[\e [\hat f_0(\bar X_{1,j}) |\bar X_{1,j}] \cdot \bar X_{1,j}\big] \nonumber\\
    &= \e \big[ f_0(\bar X_{1,j}) \cdot \bar X_{1,j} \big]. 
\end{align}
Because \eqref{eq:unbiasProof3} and \eqref{eq:unbiasProof2} are equivalent, we proved the unbiasedness that $\e[\hat {RC}(n|\bar X)] = \e_y^{\bar X \cup n}[y] - \e_y^{\bar X}[y]$, and that the noise terms do not need to be in the estimator to be unbiased.

\subsection{Proof of Proposition \ref{prop:simpT}}
\label{Apex:simpT}
First, the estimator in \eqref{eq:simpT} can be re-written as 
\begin{align}
\label{eq:simT_proof1}
     &=\frac{1}{J}\big[ \sum_{j=1}^J n_{1,j} - \sum_{j=1}^J \hat f_0(\bar X_{1,j}) \big]\cdot \bar X_1 \nonumber \\
     &=\frac{1}{J} \big[ \sum_{j=1}^J \big[\hat f_1(\bar X_{1,j},\epsilon_{1,j}) - \hat f_0(\bar X_{1,j}) \big] \big]\cdot \bar X_1.    
\end{align}
Similar to the proof in \ref{Apex:unbias}, under the assumption of (i) an additive noise model, (ii) $\e [\hat f_0(\bar X_{1,j}) |\bar X_{1,j}] = f_0(\bar X_{1,j})$, and (iii) $\e[\epsilon_{1,j}] =0$, taking the expectation and applying the law of total expectation leads to
\begin{equation}
\label{eq:simT_proof2}
    \e[\eqref{eq:simT_proof1}] =\frac{1}{J} \sum_{j=1}^J \e \big[ \big[f_1(\bar X_{1,j}) - f_0(\bar X_{1,j}) \big]\cdot \bar X_1 \big].
\end{equation}
Under the proposition assumption that the change distribution between two periods is only from an intercept, denoted as $C$, \eqref{eq:simT_proof2} becomes 
\begin{equation}
\label{eq:simT_proof3}
    \frac{1}{J} \sum_{j=1}^J \e  [C \cdot \bar X_1 \big] = C \cdot \e[\bar X_1].
\end{equation}
Following a similar strategy, Eq.\eqref{eq:unbias} can be re-written as
\begin{align}
\label{eq:simT_proof4}
     &=\frac{1}{J}\sum_{j=1}^J \big[\big( n_{1,j} -  \hat f_0(\bar X_{1,j})\big) \cdot \bar X_{1,j} \big] \nonumber \\
     &=\frac{1}{J} \sum_{j=1}^J \big[ \big(\hat f_1(\bar X_{1,j},\epsilon_{1,j}) - \hat f_0(\bar X_{1,j}) \big)\cdot \bar X_{1,j} \big].    
\end{align}
Similar to \eqref{eq:simT_proof2} (or \ref{Apex:unbias}), taking the expectation and applying the law of total expectation along with assumptions (i)-(iii) above,
\begin{align}
    \e[ \eqref{eq:simT_proof4} ] 
    &=\frac{1}{J} \sum_{j=1}^J \e \big[ (f_1(\bar X_{1,j}) - f_0(\bar X_{1,j}) \big)\cdot \bar X_{1,j} \big] \nonumber\\
    &= \frac{1}{J} \sum_{j=1}^J \e  [C \cdot \bar X_{1,j} \big] \nonumber \\
    &= C \cdot \frac{1}{J} \sum_{j=1}^J \e[\bar X_{1,j}]. \label{eq:simT_proof5}
\end{align}
Comparing \eqref{eq:simT_proof5} with \eqref{eq:simT_proof3}, now let us prove $\e[\bar X_{1}] = \e[\bar X_{1,j}]$. Assume $\e[\bar X_{1,j}] =\mu$, because
\begin{align}
\label{eq:simT_proof6}
    \bar X_1 &= \frac{\sum_j y_{1,j}}{\sum_j n_{1,j}} = \frac{\sum_j n_{1,j} \cdot \bar X_{1,j}}{\sum_j n_{1,j}} \equiv \sum_j w_j \bar X_{1,j}
\end{align}
where $w_j = \frac{n_{1,j}}{\sum_j n_{1,j}}$ and $\sum_j w_j =1$. So 
\begin{align}
\label{eq:simT_proof7}
    \e [\bar X_1] = \e [\sum_j w_j \bar X_{1,j}] &=\e[ \e [\sum_j w_j \bar X_{1,j}|{n_{1,1}, \ldots, n_{1,J}}] ] \nonumber\\
    &= \e [\sum_j w_j \e [\bar X_{1,j}] ] \nonumber\\
    &= \e [\sum_j w_j \mu ] = \e[\mu] = \mu
\end{align}
by proving $\e[\bar X_{1}] = \e[\bar X_{1,j}]$, the proof is complete.

\section{Simulation Studies Details for Section \ref{sec:Simu_Ext}}  
(note: for all the figures in this section, ``GCM-based'' denotes the GCM-DC approach, whereas ``Decomposition-based'' denotes the MTCD approach.)
\subsection{Non-rate  metric simulations}
\label{Apex:simu_noRate}

\subsubsection*{Scenario 1. Assuming causes (i.e., sibling nodes in MTCD) are independent \& causal DAG is correctly assumed}

\begin{figure}[ht]
    \centering
    \includegraphics[width=1\linewidth]{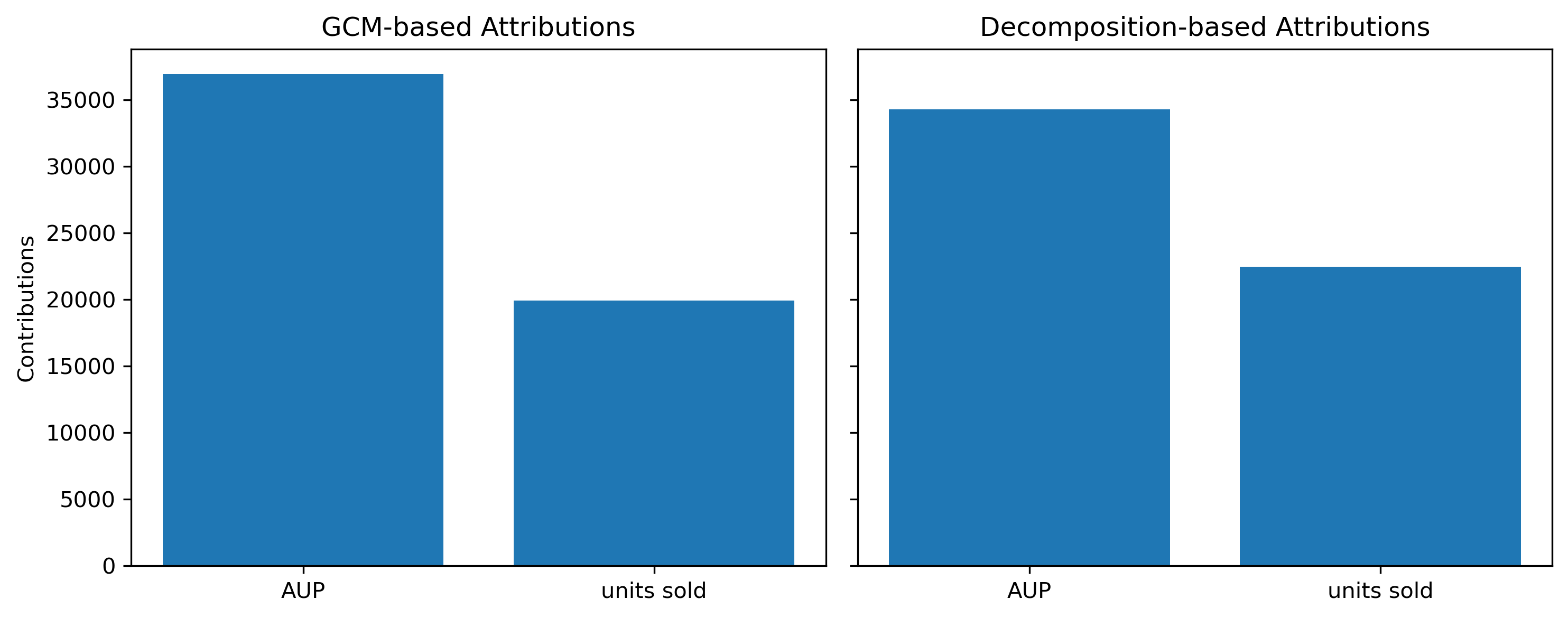}
    \caption{Scenario 1. when causal DAG is correctly assumed with independent causes}
    \label{fig:sce1}
\end{figure}

We use revenue = AUP *units sold as the simplest case. We want to test the special case when there is no relationship between units sold and AUP in simulated data, and the DAG is correctly specified with only revenue being a function of units sold and AUP, and AUP being independent of units sold. We simulated data for these three variables: {AUP, units sold, revenue} for the base period with sample size 100, and assumed that both AUP and units sold increased in mean separately in the new period with the same sample size of 100.  As discovered in the theoretical proofs section, we also scaled the contribution by sample size to make the two methods comparable and the same for all the following simulations. In this case, as observed in Fig. \ref{fig:sce1}, both approaches correctly identified the right root causes with the correct order, but with slightly different magnitudes. This difference originates exactly from the fact that GCM-DC uses Shapley, while the decomposition approach uses ordered allocation, aligned with the theoretical proofs section. 

\subsubsection*{Scenario 2a. Assuming causes (i.e., sibling nodes in MTCD) are dependent \& causal DAG is correctly assumed}

\begin{figure}[ht]
    \centering
    \includegraphics[width=1\linewidth]{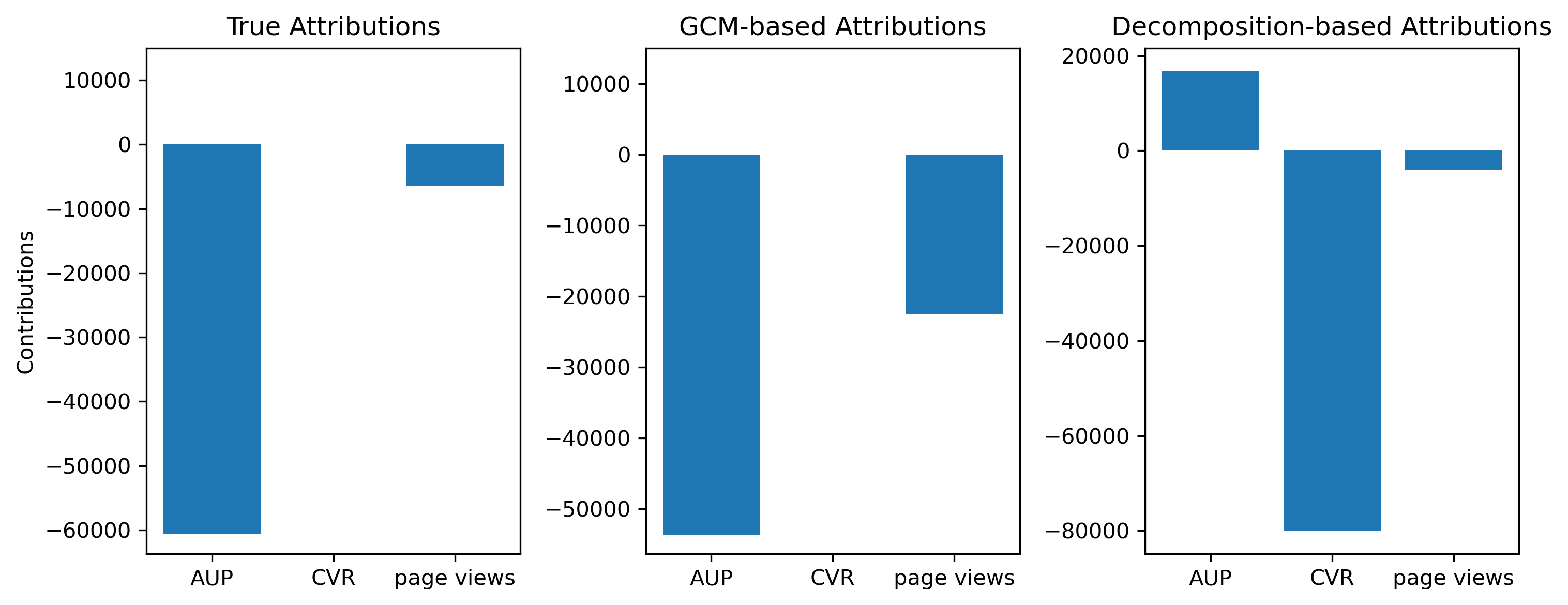}
    \caption{Scenario 2a. when causal DAG is correctly assumed (n=100): true AUP and page view effects}
    \label{fig:sce2a}
\end{figure}

Now we include one more layer of nodes under units sold and use the same 4-node causal graph shown in the real-world application in Fig. \ref{fig:4nodes}(b). We used the same DAG structure to simulate the data for one period and increase the AUP and page views in the mean for the next period. GCM-DC and MTCD (using Fig. \ref{fig:4nodes}(a)) are applied separately to find the root causes of the change in revenue. Compared with the ground truth of AUP and the page views in the leftmost panel of Fig. \ref{fig:sce2a}, GCM-DC correctly identified the true root causes with the correct order, while MTCD failed to detect the correct root causes both in existence and magnitude. This highlights the limitation of MTCD mentioned earlier that, when there is a true dependence among causes, this approach is likely to fail. It can also be seen that the magnitude of the GCM-DC contributions is slightly off from the true magnitude. We further increased the sample size from 100 to 1000 and this does not change the findings. This revealed the fact that using the Shapley value may deviate from the true contribution allocation, which will be addressed in our proposed unified approach section. 

\begin{figure}[ht]
    \centering
    \includegraphics[width=1\linewidth]{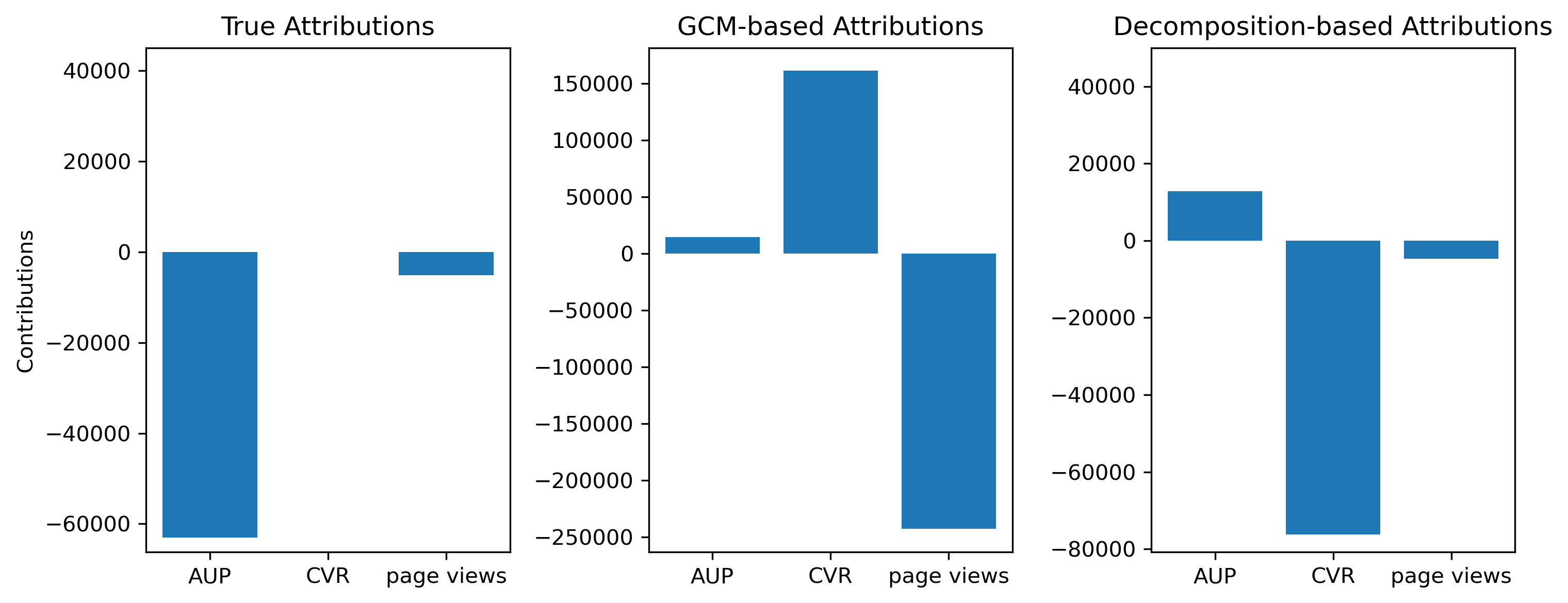}
    \caption{Scenario 2a. when causal DAG is correctly assumed (n=30): true AUP and page view effects}
    \label{fig:sce2a_n30}
\end{figure}

Separately, as noted in the GCM-DC limitations that a reasonable sample size is required to learn the conditional distribution for each children-parent pair, thus, we tested when the sample size is reduced to 30. As shown in Fig. \ref{fig:sce2a_n30}, GCM-DC no longer dectects the correct root causes, and MTCD still has similar findings to those under sample size of 100. To this end, sample size is another important factor in determining the right root causes. All following simulations are conducted with sample size 100, unless otherwise noted. 

We also examined other cases where the true root cause is AUP only and page views only, and the findings are similar (plots omitted). On the other hand, when there is only the CVR effect, it is observed and expected that both methods are able to only pick up CVR as the root cause and with a similar magnitude since there are no other factors causing the CVR to change.

\subsubsection*{Scenario 2b. weak relationship among nodes \& small signal-to-noise ratio}

\begin{figure}[ht]
    \centering
    \includegraphics[width=1\linewidth]{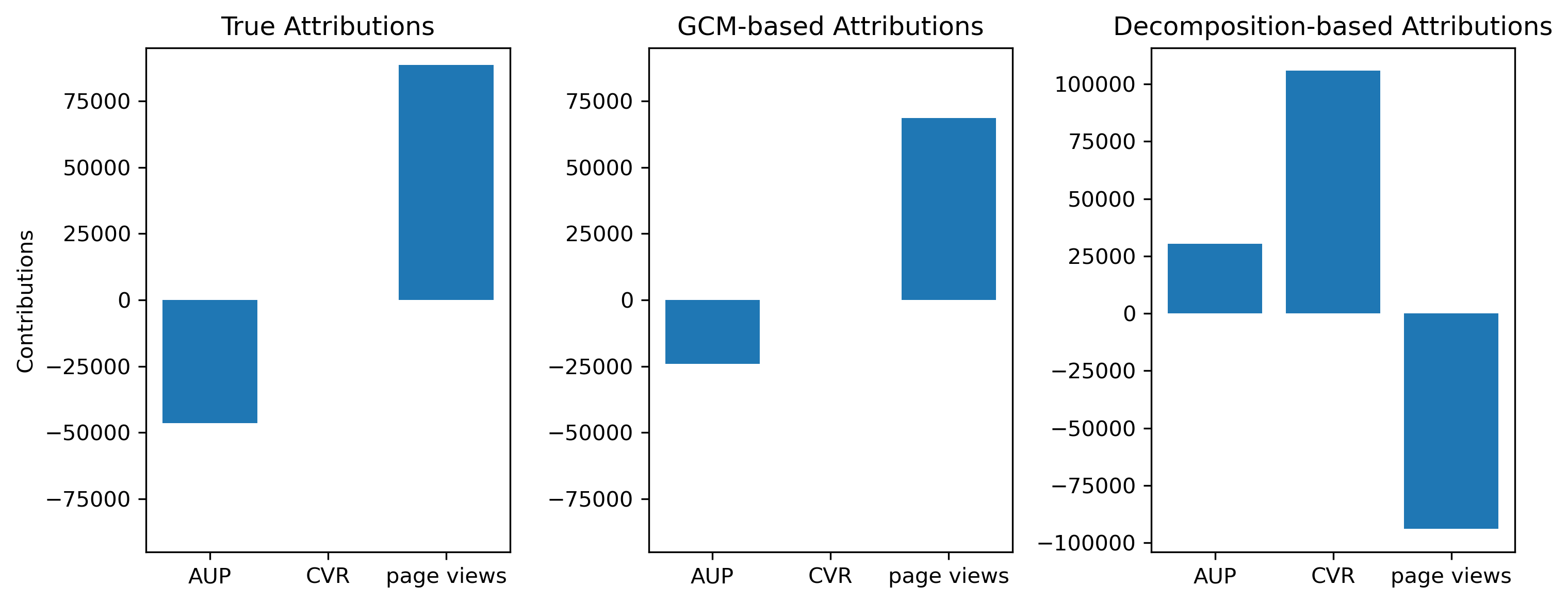}
    \caption{Scenario 2b. when causal DAG is correctly assumed with weak relations: true AUP and page view effects}
    \label{fig:sce2b}
\end{figure}

Similar to Scenario 2a, we assumed that causes are dependent \& and the causal DAG is correctly assumed; however, we manually weakened the nodes relationship by simulating page views = f(AUP) with person correlation of 0.2, and increased the variance of the noise term in CVR = f(AUP,page views) such that $R^2$ is approximately 0.2. This is designed to test whether a weak relationship between dependent nodes, and a small signal-to-noise ratio would lead to a wrong inference.  But from the simulation results in Fig. \ref{fig:sce2b}, GCM-DC continued to recognize the correct root causes with the correct order, but the magnitude is slightly off. 
This will be tackled in the proposed unified approach section. On the other hand, MTCD, as expected, was unable to identify the correct root causes due to the missing dependence between the sibling nodes (i.e., causes).

\subsubsection*{Scenario 3a. Assuming causes (i.e., sibling nodes in MTCD) are dependent but causal DAG is incorrectly assumed (i.e., same graph but wrong edge directions)}

\begin{figure}[ht]
    \centering
    \includegraphics[width=1\linewidth]{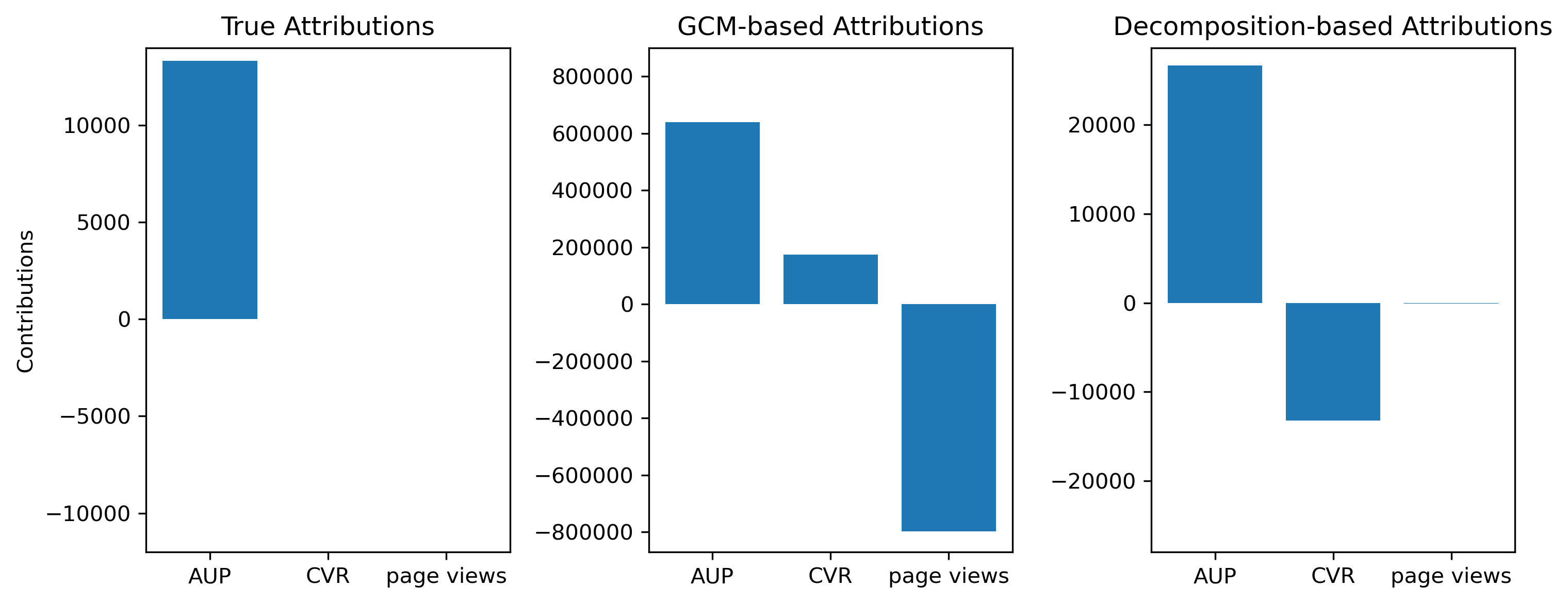}
    \caption{Scenario 3a. when causal DAG is incorrectly assumed: true AUP effect only}
    \label{fig:sce3a_AUP}
\end{figure}

We want to examine how these two methods behave if, unlike Scenario 2a, the truth causal graph is different from the assumed one. As a hypothetical example, the assumed DAG is CVR = f(AUP,page views); page views = f(AUP), but we generated data by AUP = f(CVR,page views) and page views = f(CVR) for the baseline period. In the new period, we set the true root cause to be AUP by only increasing the mean of AUP, and other nodes are changed correspondingly using the same data generating process in the baseline period. 
It is noted that in this setting, the causal graph cannot be falsified as there are edges between every possible pair, making the falsification test \cite{eulig2025falsification} impossible. From Fig. \ref{fig:sce3a_AUP}, neither method can successfully identify the correct root causes, and the GCM-DC approach detected more incorrect root causes than MTCD. 

\begin{figure}[ht]
    \centering
    \includegraphics[width=1\linewidth]{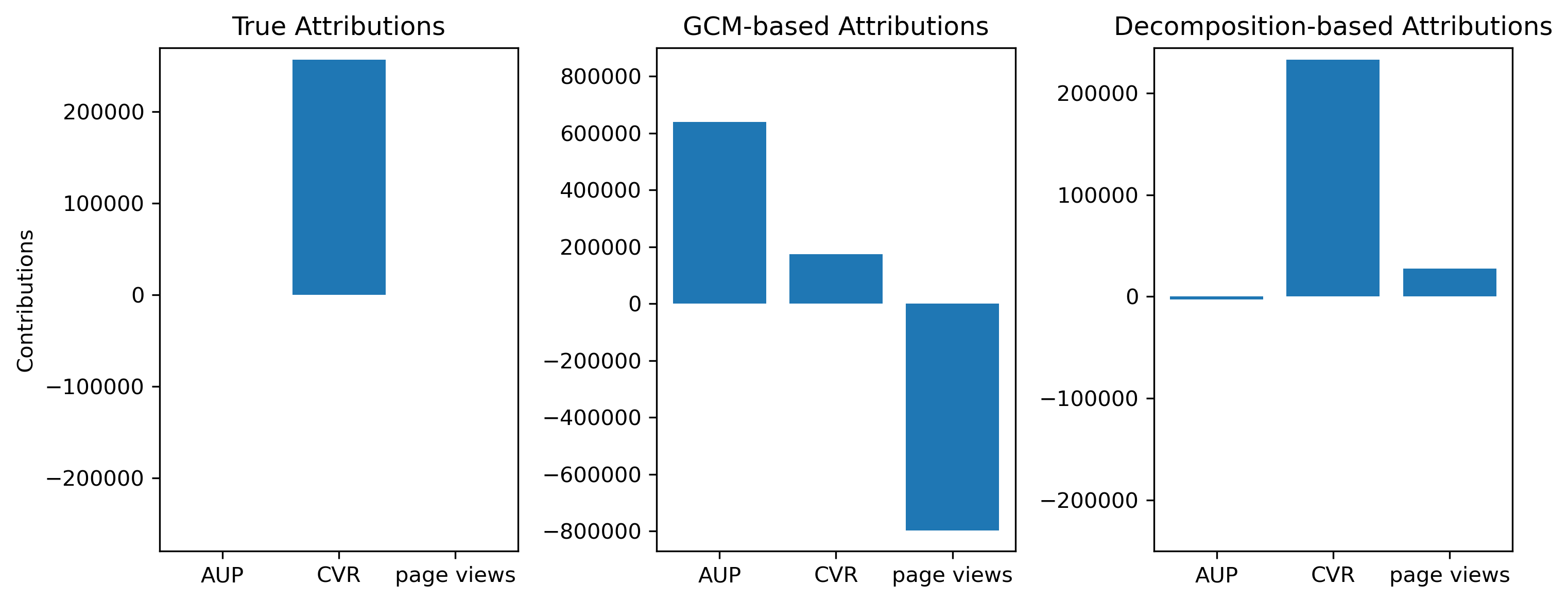}
    \caption{Scenario 3a. when causal DAG is incorrectly assumed: true CVR effect only}
    \label{fig:sce3a_CVR}
\end{figure}

When CVR is set as the only true root cause in the simulation, similarly, MTCD seemed to select fewer incorrect root causes compared to GCM-DC (Fig. \ref{fig:sce3a_CVR}). Both cases underscore the importance of correct causal DAG assumption in order for GCM-DC to have a correct inference. 

\subsubsection*{Scenario 3b. Assuming causes (i.e., sibling nodes in MTCD) nodes are dependent but causal DAG is incorrectly assumed (i.e., fewer edges)}

\begin{figure}[ht]
    \centering
    \includegraphics[width=1\linewidth]{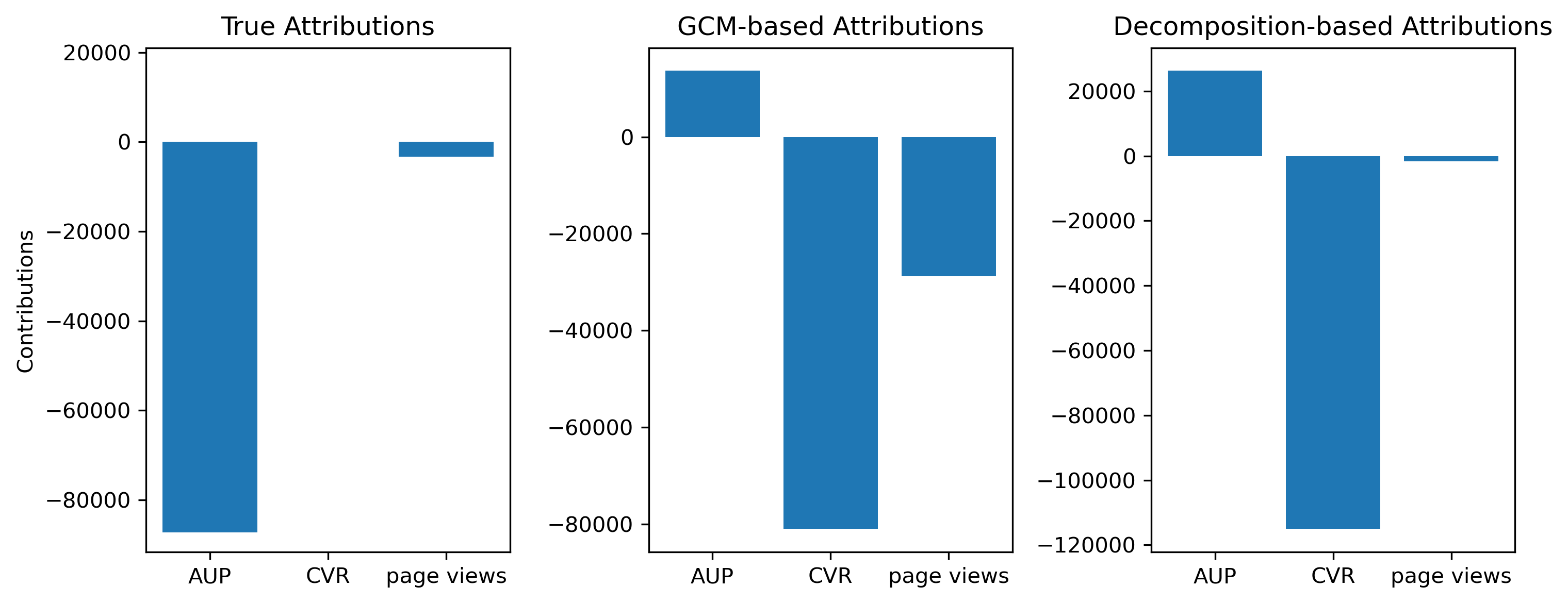}
    \caption{Scenario 3b. when causal DAG is incorrectly assumed with independence among causes}
    \label{fig:sce3b}
\end{figure}

In this scenario, we generated the data from CVR = f(AUP,page views); page views = f(AUP), but assumed a simple causal graph to be used in GCM-DC. That is, revenue is affected by all three nodes of {AUP, page views, and CVR}, but these three nodes are independent of each other. Then, we increased the mean values of AUP and page views in the process of generating the data for the new period. In summary,  both methods are somewhat similar because both have the same underlying causal structure: no dependence among causes. And the difference is mostly due to Shapley and ordered allocation. However, both failed to select the correct root causes (Fig. \ref{fig:sce3b}). In addition, the built-in falsification test in the GCM package (dowhy.gcm.falsify\_graph($\cdot$), \cite{dowhyGCM2024}) failed to reject the incorrect over-simplified DAG assumption. In such a case, the GCM-DC results become untenable. Testing in simpler true root cause cases of: (1) AUP only; and (2) page views only resulted in similar findings. When the true root cause is only CVR instead, both methods can actually recognize CVR correctly as the root cause (figure omitted here). This is simply because the other two nodes that affect CVR did not change from the previous period, and hence ignoring the dependence relationship becomes trivial. 

\subsection{Rate metric simulations}
\label{Apex:simu_Rate}
\begin{figure}[ht]
    \centering
    \includegraphics[width=0.8\linewidth]{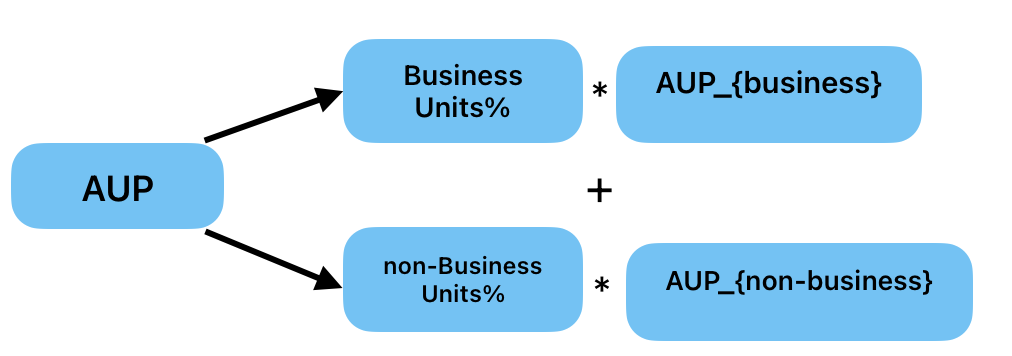}
    \caption{Rate metric decomposition}
    \label{fig:type4node}
\end{figure}

We use Fig. \ref{fig:type4node} as an illustration of the rate metric to conduct simulation studies. Namely, $\text{AUP} = \text{business units}\% \ast \text{AUP}_b + \text{non-business units}\% \ast \text{AUP}_{nb}$.

\begin{figure}[ht]
    \centering
    \includegraphics[width=0.5\linewidth]{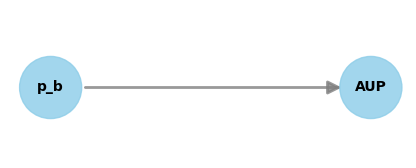}
    \caption{causal DAG for GCM}
    \label{fig:GCM_pb}
\end{figure}

\begin{figure}[ht]
    \centering
    \includegraphics[width=0.8\linewidth]{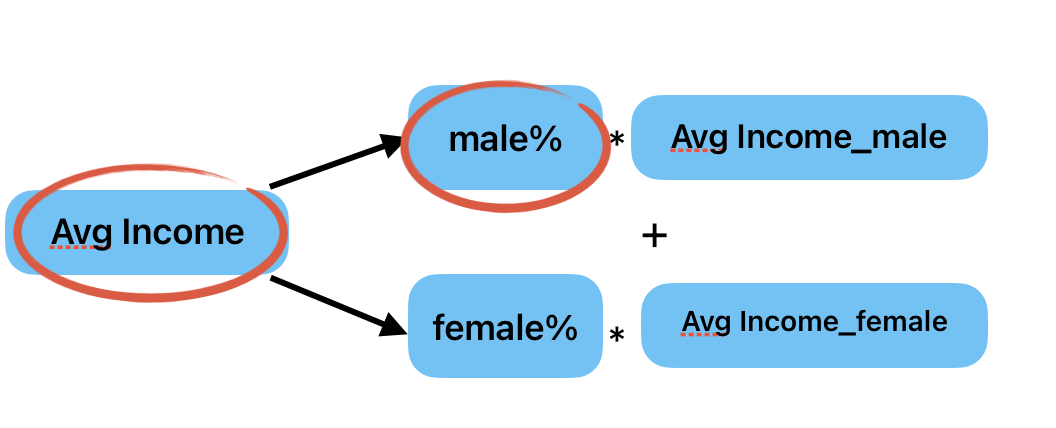}
    \caption{Analog example}
    \label{fig:analog}
\end{figure}

We want to understand how each of the 4 nodes contributes to the overall AUP change. This is not a typical causal graph, which usually involves outcome, predictor, and confounder nodes instead of including the subgroup metrics of the outcome and share nodes (i.e., business units\% and non-business units\%) given that they add up to 1. In order to produce a proper causal DAG set up for a GCM-based approach, we reframed this decomposition relationship to business units\% (denoted as $p_b$) impacting AUP as in Fig. \ref{fig:GCM_pb}. We use an analogy example to illustrate why only one node is needed instead of four.
 For example, if the outcome is average income and gender percentage per region is an observed predictor, instead of decomposing the average income in the following decomposition framework (Fig. \ref{fig:analog}), we can just model the average income against male\% instead. That is, it becomes natural that we only need to model the ones with red circles, which is typical in a regular machine learning setting.  In the linear regression case, the subgroup income for the female is equivalent to the intercept in the regression, and $\{\text{avg income}_{\text{male}} - \text{avg income}_{\text{female}}\}$is equivalent to the linear slope term. 

\subsubsection*{Scenario 4a. only $p_b$ changes}
\label{Apex:sce4a}
\begin{figure}[ht]
    \centering
    \includegraphics[width=1\linewidth]{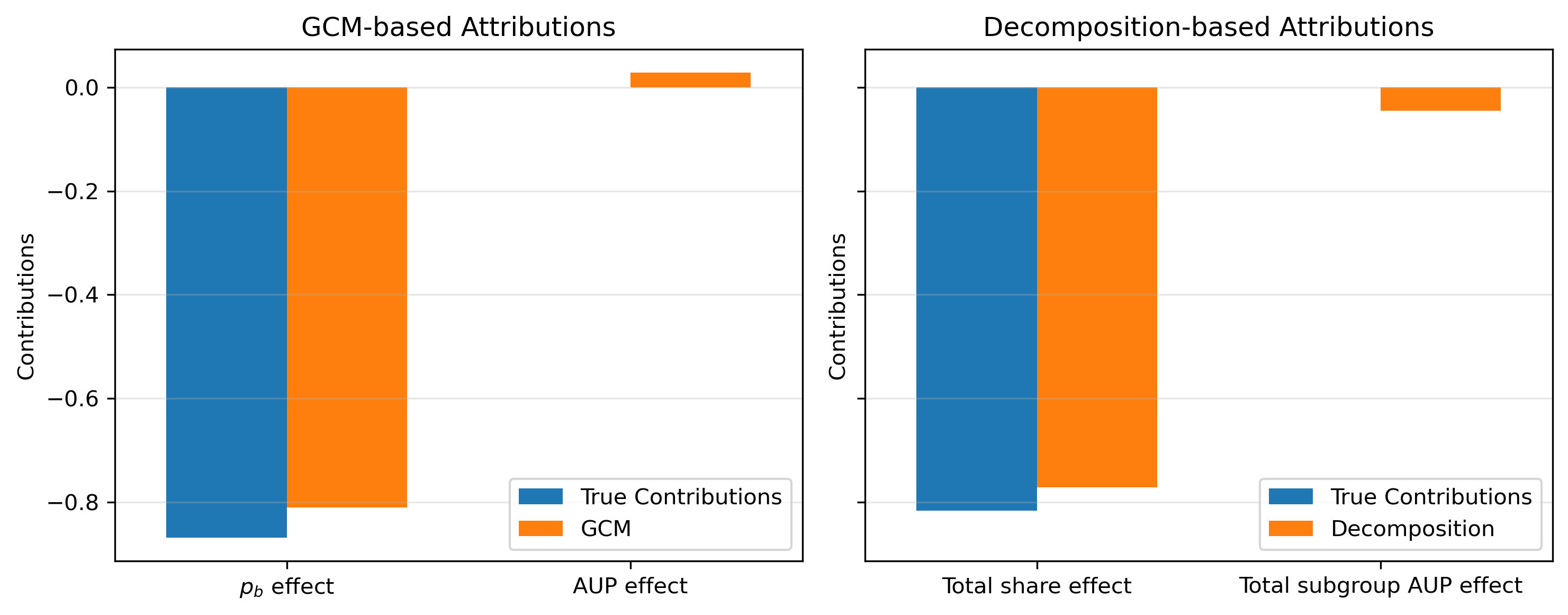}
    \caption{Scenario 4a. only $p_b$ changed}
    \label{fig:type4_case1}
\end{figure}

Returning to the AUP example with business account dimension breakdown, we simulated AUP, and units sold for business and non-business subgroups from separate normal distributions, and derived the corresponding business and non-business share percentages and the overall AUP. Then, in the new period, we simulated a case where only business units share percent ($p_b$) is the root cause of the change for the overall AUP while keeping other factors with the same distribution. Observing the results in Fig. \ref{fig:type4_case1}, firstly, according to the theoretical proofs section for the rate metric in Sec.~\ref{sec:theory_rate}, the target scale for GCM-DC and the decomposition-based approach is slightly different, leading to the difference in the true contributions not being identical in the left and right panels in Fig. \ref{fig:type4_case1}. However, both GCM-DC and the decomposition-based methods can identify the correct root cause. Secondly, while the GCM-DC uses Shapley allocation and the decomposition-based approach uses ordered allocation, but in this case of AUP distribution mechanism did not change, both allocations become equivalent. Hence, bearing the subtle difference of the target scale, and comparing the two orange bars in the left and right plots, this simulation study also verifies the special case in Sec.~\ref{sec:theory_rate} that the business node contribution of the GCM-DC is approximately equal to the sum of share nodes contribution of the decomposition-based approach.  Both methods are comparable in this case. 

\subsubsection*{Scenario 4b. both $AUP_b$ and $p_b$ change (assuming $AUP_b$ and $p_b$ are independent)}
\label{Apex:sce4b}
\begin{figure}[ht]
    \centering
    \includegraphics[width=1\linewidth]{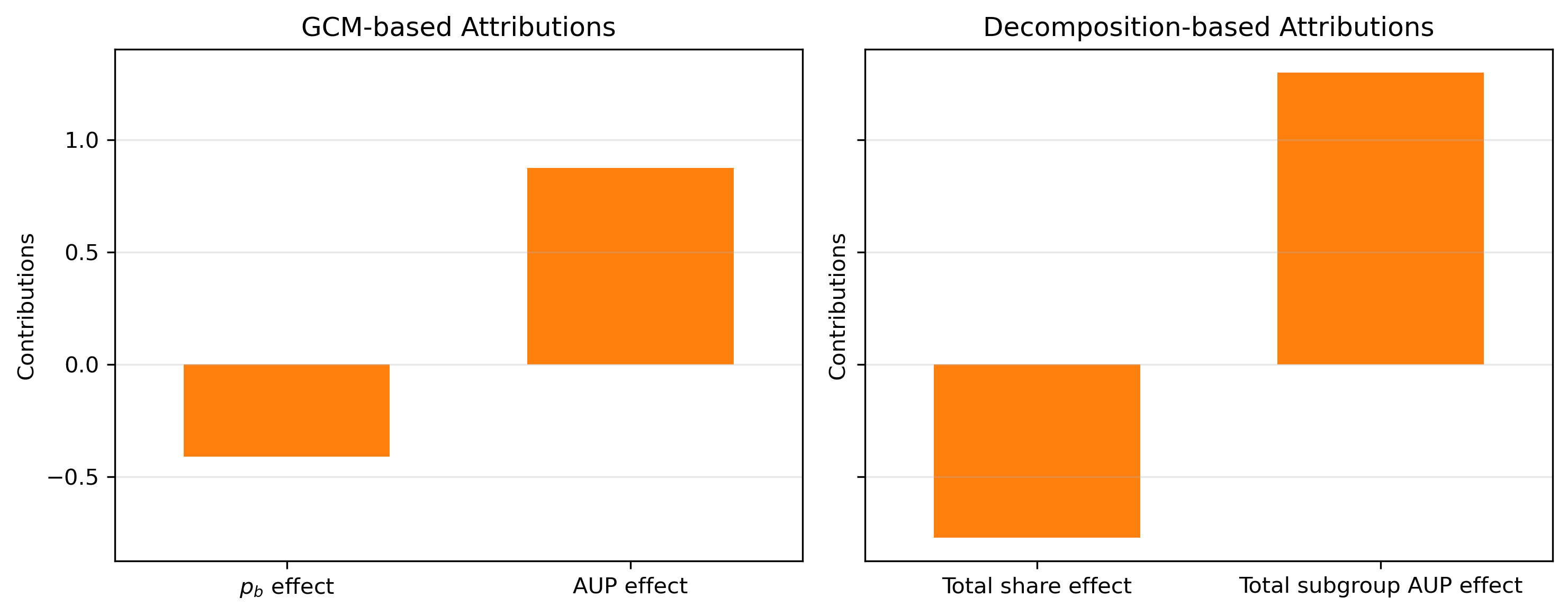}
    \caption{Scenario 4b. both $p_b$ and $\text{AUP}_b$ changed}
    \label{fig:type4_case2}
\end{figure}

In this scenario, in the new period, we not only increased the mean of units sold for business to indirectly increase the business share percent like scenario 4a, but also increased the mean of business subgroup AUP. This makes Shapley and ordered allocation no longer equivalent. In this case, even though the true increase of each variable is known, when nodes are independent, there is no unique causal ordering, hence there is no ground truth contribution of subgroup AUP and units share percentages. 
From Fig. \ref{fig:type4_case2}, both methods can correctly recognize the root causes with the right order, but the magnitude is different, and this aligns with the theoretical proofs section that the differences of the two approaches are mainly due to: (1) target scales are slightly different; and (2) the choice of Shapley vs ordered allocation. Neither allocation approach can be determined better when there is no ground truth. However, the decomposition-based approach also provides a more granular contribution of each AUP subgroup ($p_b: -0.701, p_{nb}: -0.071, AUP_b: 1.337, AUP_{nb}: -0.037$). The decomposition-based approach also successfully selected $AUP_b$ as the root cause, while GCM-DC can only know that the distribution of AUP has changed but does not know which part. 

\subsubsection*{Scenario 5: multivariate dimensions}
\label{Apex:simu_Sce5}
\begin{figure}[ht]
    \centering
    \includegraphics[width=0.9\linewidth]{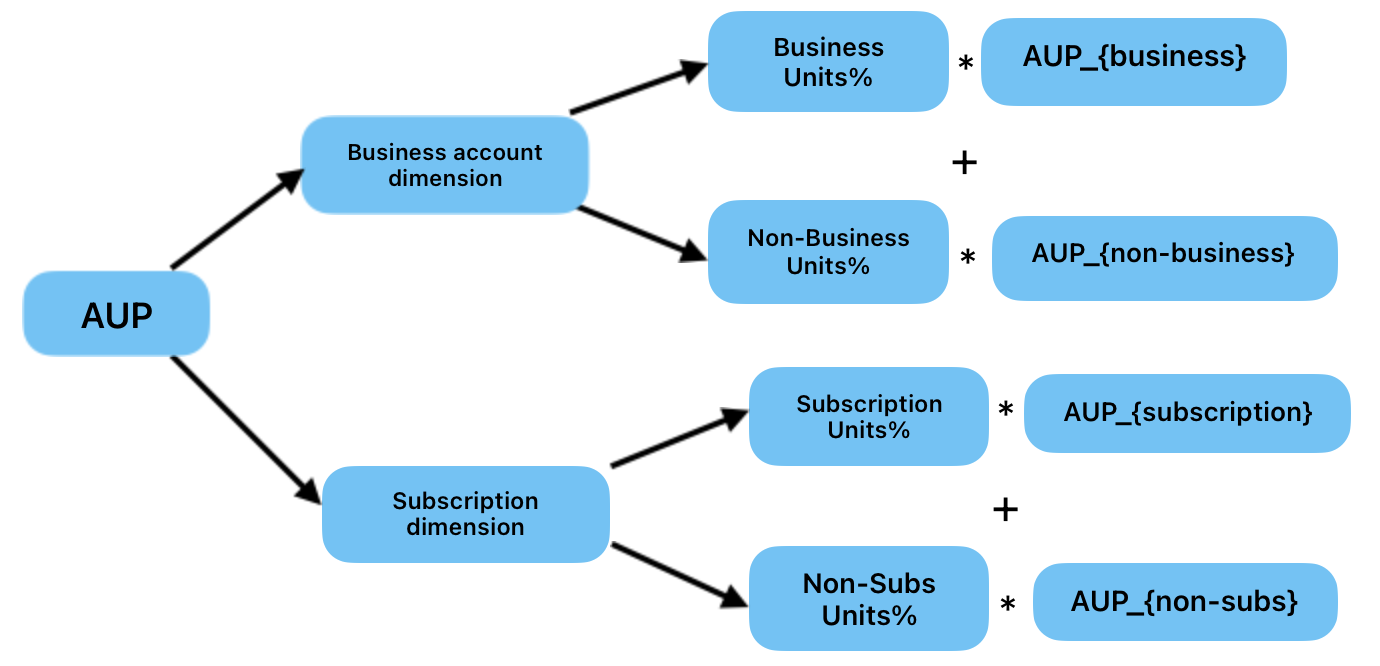}
    \caption{2-dimensional MTCD}
    \label{fig:decomp-2dim}
\end{figure}

\begin{figure}[ht]
    \centering
    \includegraphics[width=0.5\linewidth]{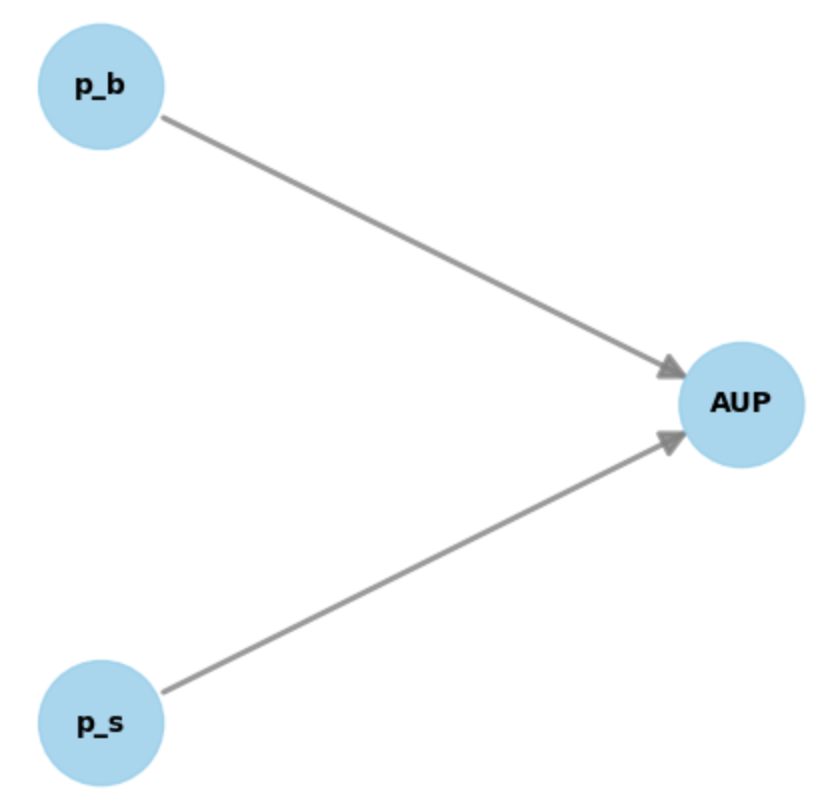}
    \caption{2-dimensional example causal DAG}
    \label{fig:pb&ps}
\end{figure}

Like the MTCD approach we developed in Section \ref{sec:tree_decomp}, it is often reasonable to break down a rate metric into multiple dimensions. Given that GCM-DC models all parent nodes together with their joint child node in each conditional distribution, while in contrast, the MTCD approach decomposes a node into multiple dimensions separately in parallel, we want to test the hypothesis whether MTCD inflates each node’s contribution compared to the GCM-DC approach. Without loss of generality, we will simulate a 2-dimensional scenario and hypothetically name them as business and subscription for easier demonstration (Fig. \ref{fig:decomp-2dim}). Similarly, we add the $p_s$ node to represent subscription’s impact in the causal graph for GCM-DC (Fig. \ref{fig:pb&ps}). We assume that the business account factor is independent from the subscription factor in this simulation. Given the model mechanism difference between GCM-DC and MTCD, the data generating process becomes more complex and was conducted as follows: Because we model
$AUP = \beta_0 + \beta_1 p_b + \beta_2 p_s + \epsilon$ for GCM-DC, and $AUP = p_b AUP_b + (1-p_b) AUP_{nb}$ for MTCD,
solving these two equations leads to: 
\begin{align}
    \e (AUP_{b}) &= \beta_0 + \beta_1 + \beta_2 \e(p_s) \label{eq:Sce5a_eq1} \\
    \e(AUP_{nb}) &= \beta_0 + \beta_2 \e(p_s) \nonumber\\
    AUP_{nb} &= (AUP - p_b AUP_b)/(1-p_b).  \label{eq:Sce5a_eq3} 
\end{align}
With these relationships, we can simulate $p_b$ and $p_s$ from two beta distributions, and pre-specify $\beta$ coefficients to generate the AUP distribution, and consequently simulate $AUP_b$ with the mean from \eqref{eq:Sce5a_eq1} above with a fixed variance, and obtain the corresponding $AUP_{nb}$ by \eqref{eq:Sce5a_eq3}. We can also solve $\e(AUP_{s})$ similarly as a function of $p_b$, and the corresponding subgroup AUP distributions for subscription can be simulated. This consitutes the data for the base period.  

\subsubsection*{Scenario 5a:  simulate only $p_b$ change}
\begin{figure}[ht]
    \centering
    \includegraphics[width=1\linewidth]{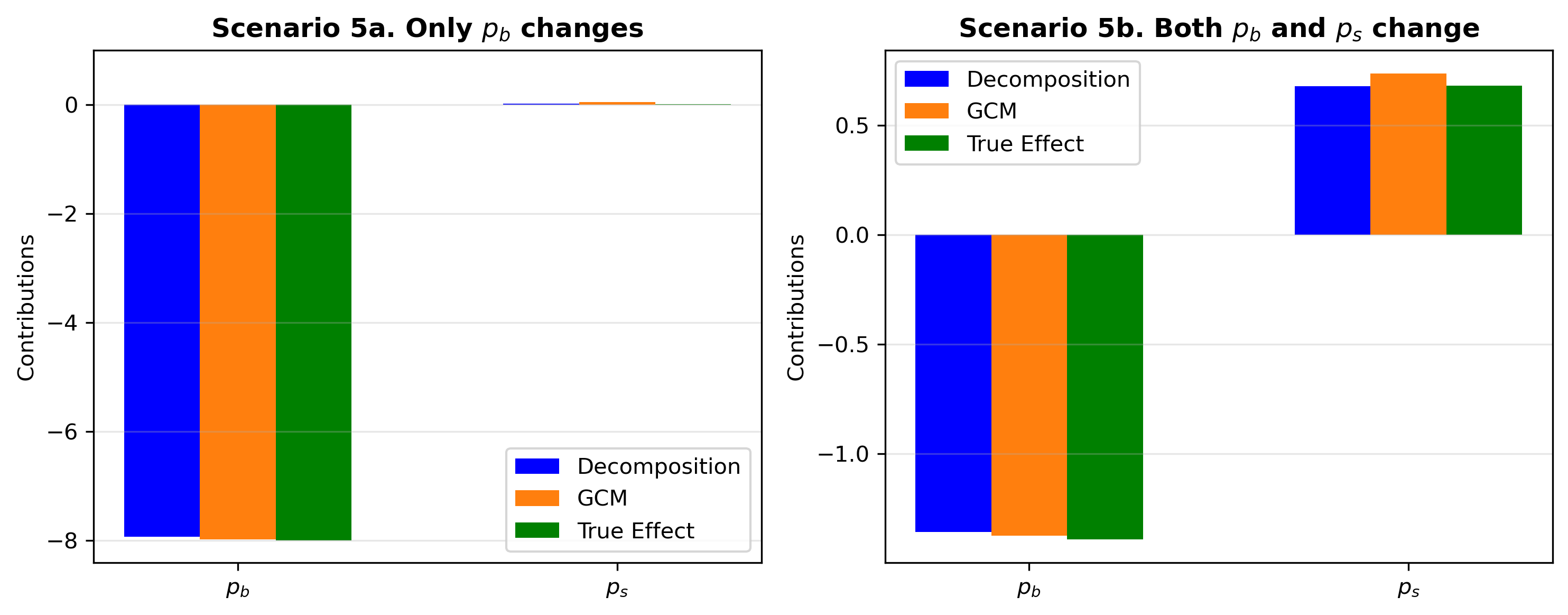}
    \caption{Scenario 5a-b. multivariate dimensions for rate metrics}
    \label{fig:type4_2dim}
\end{figure}

In the new period, if only $p_b$ is the root cause, according to the data generating process, this will consequently change the overall AUP, $AUP_s$, and $AUP_{ns}$.  We plotted the $p_b$ effect from GCM-DC, the total share effect from MTCD, and the true $p_b$ effect on the left, and similarly for subscription effect on the right within the left panel of Fig. \ref{fig:type4_2dim}. Both GCM-DC and MTCD effectively identified root causes and provided satisfactory contribution estimates. This also confirms $p_b$ from GCM-DC $\approx {p_{nb} + p_b}$ from MTCD, which aligns with the theoretical proofs section. Additionally, in MTCD, the contribution of $AUP_s + AUP_{ns}$ also approximately equals the contribution of the total share effect for business (i.e., $p_{nb} + p_b$).

\subsubsection*{Scenario 5b:  simulate $p_b$ and $p_s$ change}

When both $p_b$ and $p_s$ are the root causes, based on the data generating process, the general AUP and all subgroup AUP ($AUP_s$, $AUP_{ns}$, $AUP_b$, and $AUP_{nb}$) will be consequently changed. The right plot of Fig. \ref{fig:type4_2dim} indicates that both approaches accurately estimated the root causes in both identification and magnitude. 

In summary, under the assumption that multiple nodes are independent (e.g., $p_s$ is independent of $p_b$), computing the contributions in separate dimensions does not exaggerate the contribution to its parent node in the tree (e.g., overall AUP), proving MTCD is a valid approach.

Below are some additional data generating process details for preparing inputs needed for both GCM-DC and decomposition-based approaches throughout the simulations:
\begin{enumerate}
    \item Non-rate simulations:
We assume true mean and variance parameters to simulate the daily records, then aggregate the daily data to period-level metric through units sold, page views, and revenue in order to derive corresponding AUP, CVR on the aggregate level. These will be inputs for the decomposition-based approach. 

    \item Rate metric simulations:
We generate daily sub-group units sold and sub-group AUP under certain distributions, and then derive period-aggregated sub-group revenue, sub-group units\%, overall AUP, which are needed for the decomposition-based approach.
\end{enumerate}

\end{document}